\documentclass[12]{article}
\usepackage{amsmath,amssymb,amsthm}
\usepackage{mathtools}
\usepackage{mathrsfs}
\usepackage{bm}
\usepackage{slashed} 
\usepackage{multirow}
\usepackage{tikz}
\usepackage[utf8]{inputenc}
\usepackage{fullpage}
\usepackage{arydshln}
\usepackage{setspace}
\usepackage[margin=1in]{geometry}
\usepackage{dsfont}
\usepackage{graphicx }
\graphicspath{ {graphs/} }
\usepackage{subcaption}
\usepackage{float}
\usepackage{xcolor}
\usepackage{cite}
\usepackage{subcaption}
\usepackage{mwe}
\usepackage[toc,page]{appendix}
\usepackage{hyperref}
\usepackage{tabu}
\usepackage{physics}
\hypersetup{colorlinks=true,linkcolor=blue,anchorcolor=blue,citecolor=green,filecolor=blue
urlcolor=blue,bookmarksnumbered=true,pdfview=FitB}

\usepackage{epstopdf}

\renewcommand*{\thefootnote}{\fnsymbol{footnote}}

\begin{document}

\begin{center}
{\Large \bf \strut
Basis Light-Front Quantization for a Chiral Nucleon-Pion Lagrangian
\strut}\\
\vspace{5mm}
\today

\vspace{5mm}
{\large 
Weijie Du$^{a,b,c,}$\footnote{duweigy@gmail.com}, Yang Li$^{a,d,e}$, Xingbo Zhao$^{b,c}$, Gerald A. Miller$^{f}$ and James P. Vary{$^a$}
}  

\noindent
{\small $^a$\it Department of Physics and Astronomy, Iowa State University, Ames, Iowa 50010, USA  \\
{\small $^b$\it Institute of Modern Physics, Chinese Academy of Sciences, Lanzhou 730000, China} \\
{\small $^c$\it University of Chinese Academy of Sciences, Beijing 100049, China} \\
{\small $^d$\it Hebei Key Laboratory of Compact Fusion, Langfang 065001, China} \\
{\small $^e$\it ENN Science and Technology Development Co., Ltd., Langfang 065001, China } \\
{\small $^f$\it Department of Physics, University of Washington, Seattle, Washington 98195-1560, USA } 
}

\end{center}


\setcounter{footnote}{0} 
\renewcommand*{\thefootnote}{\arabic{footnote}}

\section*{Abstract}
We present the first application of the Basis Light-Front Quantization method to a simple chiral model of the nucleon-pion system as a relativistic bound state for the physical proton. The light-front mass-squared matrix of the nucleon-pion system is obtained within a truncated basis. The mass and the corresponding light-front wave function (LFWF) of the proton are computed by numerical diagonalization of the resulting mass-squared matrix. With the boost invariant LFWF, we calculate the probability density distribution of the pion's longitudinal momentum fraction and the Dirac form factor of the proton. 


\section{Introduction}
Developing a relativistic methodology that is broadly applicable to nuclear physics is 
important for studying high-momentum transfer experiments on nuclear targets in exclusive, nearly exclusive or inclusive processes \cite{Hen:2016kwk,Aubert:1983xm,HCheng:1987}. One of the promising methods for such investigations is the Basis Light-Front Quantization (BLFQ) approach \cite{Vary:2009gt}.

BLFQ is a non-perturbative, {\it ab initio} method, which treats relativistic quantum field theory via the Hamiltonian approach within the light-front (LF) formalism. BLFQ has been shown to be a promising tool in a range of applications \cite{Vary:2016ccz}, such as the electron anomalous magnetic moment \cite{Honkanen:2010rc,Zhao:2014xaa}, the positronium spectrum \cite{Wiecki:2014ola}, and the heavy quarkonium structure and radiative transitions \cite{Vary:2018pmv,Li:2015zda,Li:2017mlw,Tang:2018myz,Li:2018uif}. More recently, BLFQ has been applied successfully to the properties of the light mesons \cite{Jia:2018ary}, which are then extended to experiment-relevant scales by perturbative QCD evolution \cite{Lan:2019vui}. This Hamiltonian approach has also been extended to a non-perturbative scattering framework through time-dependent BLFQ (tBLFQ) \cite{Zhao:2013cma,Zhao:2013jia}.

BLFQ employs the LF formalism \cite{Dirac:1949cp,Dirac:1950pj}, where physical systems are quantized at fixed LF time $x^+=t+z$ \cite{Harindranth,Wiecki:2014ola}. The structure and dynamics of the systems are characterized by the Hamiltonian formalism. The LF vacuum has a simple structure since the Fock vacuum is an exact eigenstate of the full normal-ordered Hamiltonian \cite{Brodsky:1997de,Hiller:2016itl}. This provides access to the Fock-space expansion of the physical states in the LF field theory and thereby generates physical intuition for their underlying structures \cite{Brodsky:1997de,Hiller:2016itl}. 

BLFQ also takes the advantage of the developments in {\it ab initio} non-relativistic quantum many-body theories, such as the No-Core Shell Model (NCSM) \cite{Navratil:2000ww,Navratil:2000gs,Barrett:2013nh}, and the rapidly developing supercomputing techniques (algorithms and hardwares) (see, e.g., \cite{Vary:2018hdv} and references therein). In BLFQ, the LF mass-squared operator of a hadron system in the basis representation becomes a sparse matrix whose dimensions are controlled by truncations that respect the relativistic symmetries. By matrix diagonalization, the eigenvalues produce the mass sprectum, while the resulting eigenfunctions are the light-front wave functions (LFWFs) that encode the hadronic properties. The LFWFs can be boosted to a general Lorentz frame for calculating, e.g., form factors and scattering processes \cite{Brodsky:1997de}.

The LF quantization approach to treat a chiral model of the nucleon-pion ($N\pi$) system was first proposed by Miller \cite{Miller:1997cr,Miller:2000kv} in investigating the $N\pi$ scattering and the nucleon-nucleon scattering via perturbation theory. In this work, we will present the first non-perturbative treatment of the same chiral model via the BLFQ method. In particular, we consider a physical proton as the relativistic bound state of the $N\pi$ system. Via the BLFQ approach, we obtain the LF mass-squared matrix of the $N\pi$ system within a truncated basis. We compute the proton's mass and the corresponding LFWF by numerical diagonalization of the mass-squared matrix. Based on the LFWF, we evaluate the probability density distribution of the pion's longitudinal momentum fraction and the Dirac form factor of the proton. 

The outline of this paper is the following. We begin by introducing our adopted Lagrangian density in Sec. \ref{sec:theoryOftheChiralModel}. Then, in Sec. \ref{sec:SecTheory}, we introduce the elements of BLFQ, such as the derivation of the LF Hamiltonian density, our choice of the basis construction and truncation schemes, the derivation of the mass-squared matrix element in the basis representation, and the formalism for obtaining additional observables. In Sec. \ref{sec:ResultsDiscussions}, we present the results for the proton's mass, the proton's LFWF, the probability density distribution of the pion's longitudinal momentum fraction and the Dirac form factor of the proton. We conclude in Sec. \ref{sec:ConclusionOutlook}, where we also discuss our future plans. Some necessary mathematical details are presented in the Appendices.

\section{Theory I: the chiral model}
\label{sec:theoryOftheChiralModel}
\subsection{Lagrangian density of the chiral model}
We start with the $N\pi$ Lagrangian density (c.f., Eqs. (2.1) and (2.2) in Ref. \cite{Miller:1997cr})
\begin{align}
\mathcal{L} '_{\rm total} =&   \frac{1}{4}  { \Big( \frac{F}{g_A} \Big)^2 } {\rm Tr} \Big( \partial _{\mu} U \partial ^{\mu} U^{\dagger} \Big) + \frac{1}{4} M_{\pi}^2  { \Big( \frac{F}{g_A} \Big)^2 } {\rm Tr} \Big( U + U^{\dagger} -2 \Big) + \mathcal{L} ' _{N\pi } \ \label{eq:ChiralModelTotal}, 
\end{align}
where $\mathcal{L} ' _{N\pi } $ denotes the non-linear representation of the $N\pi$-interaction: 
\begin{align}
\mathcal{L} ' _{N\pi } = &  \bar{N} \Bigg\{ \gamma _{\mu} i \partial ^{\mu} -M_N { - } \frac{1}{1+ \big( {\pi} \big)^2 { \big(  \frac{ g_A}{ 2 F} \big) ^2 } } \Big[ \frac{g_A}{2F} \gamma _{\mu} \gamma _5 \vec{\tau} \cdot \partial ^{\mu} \vec{\pi} { { + } \frac{1}{4F^2} }  \gamma _{\mu} \vec{\tau} \cdot \vec{\pi}  \times \partial ^{\mu} \vec{\pi}   \Big] \Bigg\} N  \ \label{eq:StartingLagrangian} \ .
\end{align}
$N$ denotes the nucleon field operator. We set $F \equiv g_A \cdot f_{\pi}$ with $g_A=1.25$ being the tree-level axial-vector coupling constant and $f_{\pi}$ being the pion decay constant (set as 93 MeV in this work). $M_{\pi}$ denotes the pion mass (taken to be 137 MeV in this work). The unitary operator $U$ corresponds to the definition of the pion field (more details are available in Refs. \cite{Chang:1967zza,Miller:1997cr,Miller:2000kv}). In this work, we choose $U$ as
\begin{align}
U =& (U^{-1})^{\dagger}= \frac{1 + i \gamma _5 \vec{\tau} \cdot \vec{\pi} { \frac{g_A}{2F} } }{1 - i \gamma _5 \vec{\tau} \cdot \vec{\pi} { \frac{g_A}{2F} } }\ = \ 1 +  i \gamma _5 { \frac{g_A}{F} } \vec{\tau} \cdot \vec{\pi} - \frac{1}{2} { \Big( \frac{g_A}{F} \Big)^2 } \pi ^2 + \mathcal{O}\Big(\frac{g^3_A}{F^3} \Big)  \label{eq:U2} \ ,
\end{align}
where $\vec{\tau}$ denotes the Pauli matrices $\tau _a$ $(a=1,2,3)$, while $\vec{\pi}$ represents the pseudoscalar pion fields $\pi _{a}\ (a=1,2,3)$. 

In order to treat the chiral model [Eqs. \eqref{eq:ChiralModelTotal} and \eqref{eq:StartingLagrangian}] via the LF Hamiltonian formalism, we first manipulate the factor $\frac{1}{4F^2}$ and obtain
\begin{align}
\mathcal{L} ' _{N\pi} =& { \bar{N} \Bigg\{ \gamma _{\mu} i \partial ^{\mu} -M { - }  \frac{1}{1+ \big( {\pi} \big)^2 {  \big(  \frac{ g_A}{2F} \big) ^2 } } \Big[ \frac{g_A}{2F} \gamma _{\mu} \gamma _5 \vec{\tau} \cdot \partial ^{\mu} \vec{\pi} {  { + } \Big( \frac{g_A}{2F}  \Big)^2 } \gamma _{\mu} \vec{\tau} \cdot \vec{\pi}  \times \partial ^{\mu} \vec{\pi}   \Big] \Bigg\} N } \nonumber \\
  & { { +} \bar{N} \Bigg\{  \frac{1}{1+ \big( {\pi} \big)^2 { \big(  \frac{ g_A}{2F} \big) ^2 } } \Big[  { \frac{g^2_A -1 }{4F^2} } \gamma _{\mu} \vec{\tau} \cdot \vec{\pi}  \times \partial ^{\mu} \vec{\pi}   \Big] \Bigg\} N } \ \label{eq:oldLangrangianBeforeTransformation}.
\end{align}
We then transform/redefine the nucleon field (c.f., Refs. \cite{Miller:1997cr,Miller:2000kv}) as 
\begin{align}
N =& U ^{-\frac{1}{2}} \chi \ \label{eq:FieldRedefinition} ,
\end{align} 
where $\chi$ denotes the transformed nucleon field. The unitary operator $U ^{-\frac{1}{2}}$ is
\begin{align}
U ^{-\frac{1}{2}} =& \Big( U ^{\frac{1}{2}} \Big)^{\dagger} = \frac{1 - i \gamma _5 \vec{\tau} \cdot \vec{\pi} { \frac{g_A}{2F} } }{\sqrt{1+(\pi)^2 { \big( \frac{g_A}{2F} \big)^2 } } } \label{eq:U2dagger} .
\end{align}
The following identities hold
\begin{align}
U^{\pm \frac{1}{2}} \gamma _{\mu} =& \gamma _{\mu} U^{\mp \frac{1}{2}} \ \label{eq:U_identity1} , \\ 
i \partial ^{\mu} U^{-\frac{1}{2}} =& R^{\mu} U^{-\frac{1}{2}} \ \label{eq:U_identity2} ,
\end{align}
where we define 
\begin{align}
R^{\mu} \equiv & \frac{1}{1+ \big({\pi} \big)^2 { \big(  \frac{ g_A}{2F} \big) ^2 } } \Big[ \frac{g_A}{2F} \gamma _5 \vec{\tau} \cdot \partial ^{\mu} \vec{\pi} {  { + } \Big( \frac{g_A}{2F}  \Big)^2 } \vec{\tau} \cdot \vec{\pi}  \times \partial ^{\mu} \vec{\pi}   \Big] \label{eq:Rmu} \ .
\end{align}

Applying the transformation Eq. \eqref{eq:FieldRedefinition} and the identities Eqs. \eqref{eq:U_identity1} and \eqref{eq:U_identity2} to Eq. \eqref{eq:oldLangrangianBeforeTransformation}, we obtain the transformed $N\pi$ interaction Lagrangian density as
\begin{align}
\mathcal{L} _{N\pi} = \bar{ \chi} \Big[ \gamma _{\mu} i \partial ^{\mu} - M_N U^{{ \dagger}} \Big] \chi +  { \frac{g^2_A -1 }{4F^2} } \bar{\chi} \Bigg\{  \frac{1}{1+ \big( {\pi} \big)^2 { \big(  \frac{ g_A}{2F} \big) ^2 } } \Big[  \gamma _{\mu} \vec{\tau} \cdot \vec{\pi}  \times \partial ^{\mu} \vec{\pi}   \Big] \Bigg\} \chi
 \ \label{eq:newL_int}.
\end{align}
Note that the first term of Eq. \eqref{eq:newL_int} is of the linear representation of the chiral symmetry used by G\"ursey \cite{Chang:1967zza} and Miller \cite{Miller:1997cr,Miller:2000kv}, while the second nonlinear term is proportional to the Weinberg-Tomozawa \cite{Weinberg:1966kf,Tomozawa:1966jm} contact term. 

Overall, we obtain the transformed total $N\pi$ Lagrangian density (c.f., Eqs. (2.1) and (2.2) in Ref. \cite{Miller:1997cr}) as 
\begin{align}
\mathcal{L}_{\rm total} =&   \frac{1}{4}  { \Big( \frac{F}{g_A} \Big)^2 } {\rm Tr} \Big( \partial _{\mu} U \partial ^{\mu} U^{\dagger} \Big) + \frac{1}{4} M_{\pi}^2  { \Big( \frac{F}{g_A} \Big)^2 } {\rm Tr} \Big( U + U^{\dagger} -2 \Big) + \mathcal{L} _{N\pi}  \ , \label{eq:newTotal}
\end{align}
where $\mathcal{L} _{N\pi}$ is shown in Eq. \eqref{eq:newL_int}. Note that $\mathcal{L} '_{\rm total}$ [Eq. \eqref{eq:ChiralModelTotal}] and (after the chiral transformation Eq. \eqref{eq:FieldRedefinition}) $\mathcal{L}_{\rm total}$ are invariant when $M_{\pi} = 0 $ under the chiral transformation \cite{Miller:1997cr,Miller:2000kv}
\begin{align}
N \rightarrow & \  {\rm e} ^{i \gamma _5 \vec{\tau} \cdot \vec{a}} \ N \  ({\rm or} \ \chi \rightarrow \ {\rm e} ^{i \gamma _5 \vec{\tau} \cdot \vec{a}} \chi) \ , \\
U \rightarrow & \ {\rm e} ^{- i \gamma _5 \vec{\tau} \cdot \vec{a}} \ U \ {\rm e} ^{-i \gamma _5 \vec{\tau} \cdot \vec{a}} \ .
\end{align}

As an example, up to the terms with the two-pion processes (or, up to the order of $g^2_A/F^2$), Eq. \eqref{eq:newTotal} takes the following form:
\begin{align}
\mathcal{L}_{\rm total} = & \frac{1}{2} \partial _{\mu} \vec{\pi} \cdot \partial ^{\mu} \vec{\pi}  - \frac{1}{2} M_{\pi}^2 \vec{\pi} \cdot \vec{\pi}  + \bar{\chi} \Big[ \gamma _{\mu} i \partial ^{\mu} - M_N \Big] \chi + i { \frac{g_A}{F} } M_N \bar{\chi} \Big[ \gamma _5 \vec{\tau} \cdot \vec{\pi}  \Big] \chi  \nonumber \\
&  + \frac{1}{2} { \Big( \frac{g_A}{F} \Big)^2 } M_N \bar{\chi} \vec{\pi} \cdot \vec{\pi} \chi + { \frac{g^2_A -1 }{4F^2} } \bar{\chi}   \Big[  \gamma _{\mu} \vec{\tau} \cdot \vec{\pi}  \times \partial ^{\mu} \vec{\pi}   \Big]  \chi +  \mathcal{O} \Big(\frac{g^3_A}{F^3} \Big) \ .
\end{align}
In this initial work, we include only the terms up to the order of $\frac{g_A}{F}$ in $\mathcal{L}_{\rm total}$. The inclusion of higher order terms in $\mathcal{L}_{\rm total}$ (e.g., those proportional to ${g^2_A}/{F^2}$) will be an effort of a future work.

\section{Theory II: BLFQ approach to a chiral Lagrangian}
\label{sec:SecTheory}
In this section, we demonstrate the methodology of treating the chiral Lagrangian via the non-perturbative BLFQ approach. Some of the ideas can also be found in our recent work \cite{Du:2019qsz}. We begin by obtaining the Hamiltonian density. Then we present the details of solving the mass spectra and LFWFs. We also present the method for calculating selected observables.   
\subsection{Hamiltonian dynamics}
The dynamical $N\pi$ system can be evaluated from the eigenvalue equation
\begin{eqnarray}
P^{\mu} P_{\mu} | \Psi \rangle &=& M^2 | \Psi \rangle \ \label{eq:LFHamiltonianEquation} ,
\end{eqnarray}
where $ P^{\mu} $ is the four-vector operator of the energy-momentum. 
In the LF coordinates, the mass-squared operator, 
\begin{align}
H_{LC} \equiv P^2 = P^{\mu} P_{\mu}=P^+P^- - (P^{\perp} )^2 \ , \label{eq:MassSquared}
\end{align}
is analogous to the Hamiltonian in non-relativistic quantum mechanics. The details of the LF conventions and notations in this work can be found in Ref. \cite{Harindranth}. 
Since $P^+$ and $(P^{\perp} )^2$ are kinematical, the $P^-$, 
\begin{align}
P^- = & \frac{(P^{\perp} )^2 +M^2}{P^+} \ ,
\end{align}
is also referred to as LF Hamiltonian that generates the LF time-evolution (dynamics). $P^-$ is obtained from the Lagrangian via a Legendre transformation.  

$H_{LC}$ can be numerically evaluated in a chosen set of basis states as in BLFQ. In principle, the set of basis states has infinite dimension. In practice, one limits the basis size by introducing truncation scheme(s). The resulting finite-dimensional eigenvalue problem can be evaluated numerically as a function of cutoff(s) in the truncation scheme(s). By extrapolation to the continuum limit, the physical observables can be obtained.  Alternatively, as is frequently the case in an effective field theory, one selects a truncation to match a limiting scale in the theory.  For example, we can view the present effort as the application of an effective field theory valid on a scale below the scale where quark and gluon dynamics are essential. 

\subsection{LF Hamiltonian density by Legendre transformation}
Applying the standard Legendre transformation (see, e.g., Refs. \cite{Zhao:2013cma,Miller:1997cr}), the LF Hamiltonian density can be obtained as
\begin{align}
\mathcal{P}^- =& \underbrace{ \frac{1}{2} \partial ^{\perp} \pi _a \cdot \partial ^{\perp} \pi _a + \frac{1}{2} M_{\pi} ^2 \pi _a \pi _a  + \chi _+ ^{\dagger} \frac{(p^{\bot})^2 +M_N^2 }{p^+} \chi _+ }_{\mbox{kinetic energy for free pion and nucleon}} \nonumber \\
& + \underbrace{ \chi _+ ^{\dagger} \Big[ - \gamma ^{\bot} \cdot i \partial ^{\bot} + M_N \Big] \frac{1}{p^+} M_N \Big[ {-} i \gamma _5 { \frac{g_A}{F} } \vec{\tau} \cdot \vec{\pi} \Big] \chi _+ + \chi _+ ^{\dagger} M_N \Big[ i \gamma _5 {  \frac{g_A}{F}} \vec{\tau} \cdot \vec{\pi} \Big] \frac{1}{p^+} \Big[ \gamma ^{\perp} \cdot i \partial ^{\perp} + M_N \Big] \chi _+ }_{\mbox{one-pion emission and absorption}} \nonumber \\
& + \mathcal{O}(g_A^2/F^2)  \ \label{eq:LFHorder1f},
\end{align}
where $\chi$ denotes the nucleon field. It can be decomposed as $\chi _{\pm} = \Lambda _{\pm} \chi$, with $\Lambda _{\pm}$ being the Hermitian projection operators defined according to Eq. (A12) in Ref. \cite{Miller:1997cr}. $\chi _+$ is the dynamical component of the nucleon field. It is related to the kinematic component of the nucleon field, $\chi _- $, by the constraint equation:
\begin{align}
\chi _- &= \frac{1}{p ^+}  \gamma ^0 \Big[ \gamma ^{\perp} \cdot p ^{\perp} + M_N  \Big( { 1 { -} i \gamma _5 \frac{g_A}{F} \vec{\tau} \cdot \vec{\pi} } \Big) \Big] \chi _+ \ .
\end{align}
Note that in this prototype work that mainly focuses on demonstrating the BLFQ approach to the proton, we retain only the terms up to the order of $g_A/F$ as for the interaction terms, which correspond to the processes of single-pion emission/absorption. Higher-order terms, such as the $\pi ^2$ terms (c.f., Refs. \cite{Miller:1997cr,Miller:2000kv}), are expected to be corrections to the current calculation and will be the topic of a future work. 

\subsection{Basis representation: construction and truncation schemes}
\subsubsection{Symmetries}
The BLFQ methodology of constructing the basis for carrying out the matrix eigenvalue solution of the LF mass-squared operator $H_{\rm LC}$ is discussed in Refs. \cite{Vary:2009gt,Zhao:2013cma,Wiecki:2014ola}. In particular, we need to pay specific attention to the symmetries of the LF Hamiltonian $P^-$. These symmetries are: (1) the translational symmetry in the longitudinal direction, which results in the conservation of the total longitudinal momentum $P^+$; (2) the rotational symmetry in the transverse direction, which means that the projection of the total angular momentum is conserved; (3) the conservation of net fermion number; and (4) the transverse boost invariance. In this work, we take the neutron and proton masses to be the same ($=M_N$) and also take the masses of the charged and neutral pions to be the same ($=M_{\pi}$). By doing this, the $\mathcal{L}_{\rm total}$ respects isospin symmetry and the isospin projection of the constituent system is conserved. We construct the LF basis set according to these symmetries.

\subsubsection{Single-particle basis}
\label{sec:SpBasis}
We start with constructing the single-particle (s.p.) basis. In the longitudinal direction, we employ the discretized plane wave basis $\{ | p ^ + \rangle \}$. In particular, we constrain a particle in a longitudinal box of length $x_+=L$ and apply the periodic (anti-periodic) boundary condition to the boson (fermion). The longitudinal momentum is discretized as
\begin{align}
p^+ = \frac{2 \pi}{L} j \ \label{eq:singleParticleRelation1}, 
\end{align}
with $j = 1, 2, 3, \cdots$ for the boson and $j= \frac{1}{2}, \frac{3}{2}, \frac{5}{2}, \cdots$ for the fermion. Note that we exclude the ``zero modes" ($j=0$) for the bosons (pions in this work). The purpose of neglecting such zero modes is to avoid introducing a counterterm that would be required to manage the divergence at $p^+=0$ arising from the kinetic energy term in our LF Hamiltonian (see, e.g., Appendix \ref{sec:ContributionToLFHamiltonian}).

It is useful to define the longitudinal momentum fraction $x$ in terms of the total longitudinal momentum $P^+$ as the Bjorken variable
\begin{align}
x \equiv \frac{p^+}{P^+} \ = \ \frac{j}{K} \ \label{eq:singleParticleRelation2} ,
\end{align}
where the dimensionless parameter $K$ is related to $P^+$ via the relation $P^+ = \frac{2\pi}{L}K$. 

In the transverse direction, we employ the two dimensional harmonic oscillator (2DHO) basis. As explained in the Appendix \ref{sec:2DHObasis}, the 2DHO basis in the momentum representation can be labeled by the radial number $n$ and the angular quantum number $m$. Adopting the 2DHO basis in the transverse direction provides us with means to insure the transverse boost invariance of the LF kinematics \cite{Vary:2009gt,Brodsky:1997de},  as discussed further in Sec. \ref{sec:COMfactorization} below. 

In addition to the momentum space, we also have the the spin and isospin degrees of freedom for the $N\pi$ model. The s.p. basis can thus be classified according to the following set of quantum numbers:
\begin{align}
| \alpha \rangle = & | x, n, m, s, t \rangle \ \label{eq:labelOfQuantumNumbers} ,
\end{align}
where $s$ denotes the helicity and $t$ denotes the projection of the isospin of the particle. It is understood that the nucleons are of spin $\frac{1}{2}$ and isospin $\frac{1}{2}$, while pions are of spin $0$ and isospin $1$. The orthonormality relation of the s.p. basis is 
\begin{align}
\langle x, n, m, s, t | x', n', m', s', t' \rangle = \delta _{x,x'} \delta _{n,n'} \delta _{m,m'} \delta _{s,s'} \delta _{t,t'} \ \label{eq:basisOrthoNomalty}.
\end{align}

Note that we present the form of the s.p. basis Eq. \eqref{eq:labelOfQuantumNumbers}, as well as its orthonormality relation Eq. \eqref{eq:basisOrthoNomalty}, for the brevity in the notation/discussion in Sec's. \ref{sec:SpBasis} and \ref{sec:SymmtryAndBasisSelection}. In practice, we take the s.p. basis for the pion field as
\begin{align}
|\alpha \rangle =& | x, n, m, \lambda \rangle \ ,
\end{align}
where we omit the unneeded helicity label for the pion field and denote its isospin projection as $ \lambda $ for clarity (see, e.g., Eq. \eqref{eq:twoParticleBasis}). The nucleon field basis still bears the form of Eq. \eqref{eq:labelOfQuantumNumbers}.

\subsubsection{Multi-particle basis}
\label{sec:SymmtryAndBasisSelection}
The multi-particle basis is constructed as a direct product of the s.p. bases ($ \otimes | \alpha \rangle $). According to the symmetries of $P^-$ for the $N\pi$ system, we require the quantum numbers for all the constituent particles (labeled by $i$) in the retained multi-particle basis states to satisfy the following relations
\begin{align}
\sum _i p^+_i = P^+,  \ \sum _i m_i + \sum _i s_i = M_J, \ \sum _i t_i =T_z,\ \sum _i n_f^i = N_f \ \label{eq:SymmetryIdentitiesForBasis}.
\end{align}
The first identity requires all the basis states to have the same total longitudinal momentum. It is equivalent to 
\begin{align}
\sum _i j_i = K \ {\rm or} \  \sum _i x_i = 1 \ , 
\end{align}
according to Eqs. \eqref{eq:singleParticleRelation1} and \eqref{eq:singleParticleRelation2} for the fixed box-length $L$ and the total longitudinal momentum $P^+$. The second identity in Eq. \eqref{eq:SymmetryIdentitiesForBasis} states the conservation of the projection of the total angular momentum $M_J$, which is produced by the helicity $s$ and the projection of the orbital angular momentum $m$ of each constituent particle. (Note that the total angular momentum $J$ is, however, not a good quantum number in the LF basis states.) The third identity in Eq. \eqref{eq:SymmetryIdentitiesForBasis} states that the projection of the total isospin $T_z$  or, equivalently, total charge of the system is conserved.  The last identity in Eq. \eqref{eq:SymmetryIdentitiesForBasis} refers to the conservation of the net fermion number $N_f$, where $n_f^i=1$ for a nucleon and $n_f^i=0$ for each pion.

\subsubsection{Truncation scheme}
We apply three truncations in this work. First, the number of Fock sectors for the $N\pi$ system is truncated at the nucleon plus one-pion sector
\begin{eqnarray}
| N _{\rm phys} \rangle  &=& a |N \rangle + b | N \pi \rangle \ \label{eq:fockSectorTruncation},
\end{eqnarray}
with the amplitudes $a= \langle N | N _{\rm phys} \rangle $ and $b = \langle N \pi | N _{\rm phys} \rangle $. It is also possible to include higher Fock sectors, e.g., $| N \pi \pi \rangle $. However, we would postpone this to future work. According to the Fock sector truncation Eq. \eqref{eq:fockSectorTruncation}, we have the net fermion number $N_f=1$ for all the basis states.

According to the Fock sector truncation Eq. \eqref{eq:fockSectorTruncation}, the LF basis set in this work is
\begin{align}
\{ | \xi \rangle \} = \{ | \xi _N \rangle \} \oplus \{ | \xi_{N\pi} \rangle \} \ \label{eq:LihntFrontbasisSet}. 
\end{align}
For the $|N\rangle$ sector, the LF basis set is 
\begin{align}
 \{ | \xi _N \rangle \} = \{ | x_N , n_N, m_N, s_N, t_N \rangle \} \ ,
\end{align} 
with $x_N =1$ due to the conservation of the longitudinal momentum. For the $| N \pi \rangle $ sector, the LF basis set is  
\begin{align}
\{ | \xi_{N\pi} \rangle \} = \{ | x_N, n_N, m_N, s_N, t_N; x_{\pi} , n_{\pi} , m_{\pi} ,  \lambda \rangle \} \ \label{eq:twoParticleBasis}.
\end{align}
Note that $x_{\pi} \neq 0$ since we exclude the zero mode of the pion field in the longitudinal direction. Due to the conservation of the total longitudinal momentum, we also have $x_N + x_{\pi}=1$ and $0<x_N <1$ for the $| N \pi \rangle $ sector.

Second, we cut off the total longitudinal momentum for the many-body basis state as
\begin{align}
K = K_{\rm max}  ,
\end{align}
which makes the number of the longitudinal modes finite \cite{Hornbostel:1988fb}. The longitudinal continuum limit can be approached at the limit of $K _{\rm max} \rightarrow  \infty $ for a given box length $L$. 

Third, we truncate the number of the modes in the transverse direction for the many-body basis states by restricting the number of maximal excitation quanta, $N_{\rm max}$, as
\begin{align}
\sum _i (2n_i + |m_i| +1) \leq N_{\rm max} \label{eq:NmaxTruncation} ,
\end{align}
where $i$ denotes the constituent particles. By taking $N_{\rm max} \rightarrow  \infty $, the continuum limit in the transverse direction is realized.

\subsubsection{UV and IR cutoffs}
The 2DHO basis parameters are related, intrinsically, to the ultraviolet (UV) and infrared (IR) cutoffs of the model space \cite{Coon:2012ab,Furnstahl:2012qg}. In the momentum space, the UV and IR cutoffs can be, respectively, approximated by the basis truncation parameter $N_{\rm max}$ and the basis strength $b$ as
\begin{align}
p^{\perp}_{\rm max} \approx & b \sqrt{2 N_{\rm max}} \ , \label{eq:UVcut} \\
p^{\perp}_{\rm min}  \approx &  b/\sqrt{2 N_{\rm max}} \label{eq:IRcut} \ .
\end{align}

\subsubsection{Factorization}
The application of the 2DHO s.p. basis in the transverse direction with $N_{\rm max}$ truncation admits an exact factorization of the LFWF into ``intrinsic" and ``center of mass" (CM) components \cite{Wiecki:2014ola, Caprio:2012rv, Barrett:2013nh,Li:2013cga}. Taking advantage of this factorization, the spurious CM excitation due to the adoption of the 2DHO s.p. basis can be eliminated by the use of a Lagrange multiplier term as explained below (Sec. \ref{sec:COMfactorization}). The analogous factorization scheme has been adopted in the studies of nuclear structure (c.f., Ref. \cite{Caprio:2012rv,Barrett:2013nh}), where the three dimensional harmonic oscillator basis is adopted.

\subsection{Mode expansions}
The pion field can be expressed in terms of the creation and annihilation operators \cite{Zhao:2013cma,Wiecki:2014ola}
\begin{align}
\pi _a( x) = & \sum _{k ^+} \sum _{\lambda = -1}^{\lambda =1}  \frac{1}{ \sqrt{2Lk^+}} \int \frac{d^2 k^{\perp}}{(2\pi)^2} \Big[ a(k, \lambda) \varepsilon _a (\lambda) e^{-ikx} + a^{\dagger} (k,\lambda) {\varepsilon _a}^{\ast} (\lambda) e^{ikx} \Big] \ \label{eq:pionModeExpansion},
\end{align}
where we make it explicit that we are discretizing the longitudinal momenta and we introduce the following polarization vectors for the isospin degree of freedom of the pseudoscalar pion field $\pi _a \ (a=1,2,3)$
\begin{align}
\varepsilon (+1) = \frac{1}{\sqrt{2}} (1,i,0)^T , \ 
\varepsilon (0) = (0,0,1)^T , \ 
\varepsilon (-1) = \frac{1}{\sqrt{2}} (1,-i,0)^T \ ,
\end{align}
with $\varepsilon ^{\dagger}(\lambda _i)\varepsilon (\lambda _j) = \delta _{\lambda _i, \lambda _j}$ and $\varepsilon (-\lambda) = {\varepsilon} ^{\ast} (\lambda)$. The subscript ``a" also indicates the component of the polarization vector $\varepsilon (\lambda)$. $\lambda$ denotes the projection of the isospin of the physical pions, i.e., $\pi ^{\pm}$ and $\pi ^0$. The commutation relation, in terms of the discretized longitudinal momentum [Eq. \eqref{eq:singleParticleRelation2}], is
\begin{align}
[a(k, \lambda), a^{\dagger}(k', \lambda ')] =& (2\pi)^2 \delta ^{(2)}(k_{\perp}-k'_{\perp}) \delta _{\lambda , \lambda '} \delta _{k^+,{k'}^+} \ .
\end{align}

Similar to the pion field, the nucleon field can be represented with the creation and annihilation operators 
\begin{align}
\chi _+ (x) =& \sum _{p^+} \sum _{s ,t} \frac{1}{ \sqrt{2L}} \zeta (s) T(t) \int \frac{d^2 p^{\perp}}{(2\pi)^2} \Bigg[ b(p,s,t) e^{-ipx} + d^{\dagger}(p,-s,-t) e^{ipx} \Bigg] \ \label{eq:nucleonModeExpansion},
\end{align}
where
\begin{align}
\zeta (+\frac{1}{2}) = (1,0,0,0)^T,\ \zeta (-\frac{1}{2}) = (0,1,0,0)^T \ , \\
T(+\frac{1}{2}) = (1,0)^T, \ T(-\frac{1}{2}) = (0,1)^T \ .
\end{align}
With the discretized longitudinal momentum [Eq. \eqref{eq:singleParticleRelation2}], the anticommutation relations are
\begin{align}
\{ b(p,s,t), b^{\dagger}(p',s',t') \} =& (2\pi)^2 \delta ^{(2)}(p_{\perp}-p'_{\perp}) \delta _{s ,s '} \delta _{t,t'} \delta _{p^+,{p'}^+} \ , \\
\{ d(p,s,t), d^{\dagger}(p',s',t') \} =& (2\pi)^2 \delta ^{(2)}(p_{\perp}-p'_{\perp}) \delta _{s ,s '} \delta _{t,t'} \delta _{p^+,{p'}^+} \ .
\end{align}
Note that with our limited Fock space [Eq. \eqref{eq:fockSectorTruncation}], the independent field for the anti-nucleon is not included. 

The anti/commutation relations for the equal light-front time fields are
\begin{align}
[\pi _a (x), \pi _b (y)]_{x^+=y^+} =& - \frac{i}{4} \epsilon (x^- - y^-) \delta ^{(2)} (x^{\perp}-y^{\perp}) \delta _{ab} \ , \\
\{ \chi _+ (x) , \chi _+ ^{\dagger} (y) \}_{x^+=y^+} =& \frac{1}{2}\gamma ^0 \gamma ^+ \delta (x^- - y^-) \delta^{(2)} (x^{\perp} - y^{\perp}) \ .
\end{align}
$\epsilon (x) = \theta (x) - \theta (-x)$ is the antisymmetric step function, where the step function is
\begin{align}
\theta (x) = 0 \ \ {\rm for} \ x \leq 0 \ ; \ \theta (x) = 1 \ \ {\rm for} \ x>0 \ .
\end{align}
The relations $\frac{\partial \epsilon (x)}{\partial x} = 2 \delta (x)$ and $|x|=x \epsilon (x) $ hold. For the representation of the gamma matrices in this work, we follow the convention in Ref. \cite{Harindranth}.

The creation and annihilation operators in terms of the 2DHO basis with the momentum fraction weighted variables (see definitions in Appendix \ref{sec:2DHObasis}) are
\begin{align}
a(x, k^{\perp},\lambda) =& \frac{1}{\sqrt{x}} \sum _{n,m} \Psi _n^m \Big(\frac{k^{\perp}}{\sqrt{x}} \Big) \alpha (x, n,m, \lambda) \ \label{eq:aToAlpha}, \\ 
b(x, p^{\perp}, s ,t) = & \frac{1}{\sqrt{x}} \sum _{n,m} \Psi _{n}^m \Big(\frac{q^{\perp}}{\sqrt{x}} \Big) \beta (x, n,m,s ,t) \ \label{eq:bToBeta},
\end{align}
with the anti/commutation relations
\begin{align}
[\alpha (x,n,m,\lambda), \alpha ^{\dagger}(x',n',m',\lambda)] = & \delta _{x,x'} \delta _{n,n'} \delta _{m,m'} \delta _{\lambda , \lambda '} \ , \\
\{ \beta (x,n,m,s ,t) , \beta ^{\dagger} (x',n',m',s ',t')  \} =&  \delta _{x,x'}  \delta _{n,n'} \delta _{m,m'} \delta _{s , s '} \delta _{t,t'} \ .
\end{align}

\subsection{Mass-squared operator}
\label{sec:COMfactorization}
The adoption of the 2DHO s.p. basis in the transverse direction allows spurious CM excitation for the mass spectrum. In order to eliminate the states with spurious CM excitation in the BLFQ approach, we follow Ref. \cite{Wiecki:2014ola} and introduce a Lipkin-Lawson Lagrange multiplier term \cite{Gloeckner:1974sst,Lipkin:1958zza} to the mass-squared operator $H_{\rm LC}$ [Eq. \eqref{eq:MassSquared}]. The modified mass-squared operator is  
\begin{align}
H =&  \ H_{\rm LC} + \Lambda (H_{\rm CM} - 2b^2 I) \  \label{eq:HwithCMregulated},
\end{align}
where $\Lambda >0 $ is the Lagrangian multiplier. The intrinsic motion in the solutions is not influenced by this Lawson term $(H_{\rm CM} - 2b^2 I)$ due to the factorization of the LFWF in the 2DHO basis with $N_{\rm max}$ truncation. The mass spectrum of the intrinsic motion is only determined by the intrinsic part of the LFWF below the scale $2\Lambda b^2$. The CM motion is governed by
\begin{align}
H_{\rm CM} = \big( P^{\perp} \big)^2 + b ^4 \big( R^{\perp} \big)^2 ,
\end{align}
where the CM momentum and coordinate in the transverse direction are respectively
\begin{align}
P^{\perp} = \sum _i p^{\perp} _i , \ R^{\perp} = \sum _i x_i r^{\perp} _i .
\end{align}
In terms of momentum fraction weighted variables, these CM variables are
\begin{align}
P^{\perp} = \sum _i \sqrt{x_i} q^{\perp} _i , \ R^{\perp} = \sum _i \sqrt{ x_i} s^{\perp} _i .
\end{align}
$H_{\rm CM}$ satisfies the eigenequation
\begin{align}
H_{\rm CM} |\tilde{N} \tilde{M} \rangle =  (2 \tilde{N} +| \tilde{M} | +1) 2 b^2 |\tilde{N} \tilde{M} \rangle \label{eq:COMSpectrum} ,
\end{align}
where $| \tilde{N} \tilde{M} \rangle $ is the eigenvector that corresponds to the eigenvalue $\mathcal{E} _{\tilde{N} \tilde{M}} = (2\tilde{N} +| \tilde{M} | +1) 2 b^2$. Based on Eq. \eqref{eq:COMSpectrum}, it is easy to see that  the states with CM excitation (i.e., states with $ \tilde{N} \neq 0$ and/or $ \tilde{M} \neq 0$) are lifted in the spectrum; only the states with the lowest CM mode (i.e., states with $ \tilde{N} = \tilde{M} =0$) remain without a shift \cite{Maris:2013qma}. In general, the spectrum of $H$ is a set of equally spaced approximate copies \footnote {The copies are not exact numerical copies since the addition of available quanta to the CM motion means the loss of available quanta in the relative motion.} (named as subspectra), with the spacing characterized by $2 \Lambda b^2$ for every additional excitation quanta in the CM degree of freedom. In practice, we choose $\Lambda$ to be sufficiently large such that the subspectra with different CM modes are well separated. 

Making use of the LF Hamiltonian density $\mathcal{P} ^-$ [Eq. \eqref{eq:LFHorder1f}] and the mode expansions for the pion and nucleon fields [Eqs. \eqref{eq:pionModeExpansion} and \eqref{eq:nucleonModeExpansion}], we calculate the mass-squared operator [Eq. \eqref{eq:MassSquared}] as 
\begin{align}
H_{\rm LC} = & P^+ \underbrace{ \big( P^-_{\rm KE_N} + P^-_{\rm KE_{\pi}} { - } P^-_{\rm int} \big)}_{P^-} - \big(P^{\perp} \big)^2 \label{eq:HLCoperator},
\end{align}
where $P^-_{\rm KE_N}$ and $P^-_{\rm KE_{\pi}}$ denote the contributions from a free nucleon and a free pion, respectively. $P^-_{\rm int}$ denotes the $N\pi$-interaction term (only for one-pion processes) in this work. The detailed expressions of $P^+P^-_{\rm KE_N}$, $P^+P^-_{\rm KE_{\pi}}$ and $P^+P^-_{\rm int}$ are shown in Appendix \ref{sec:ContributionToLFHamiltonian}.

\subsection{Observables}
\label{sec:observableSec}
In terms of the LF basis set $\{ | \xi \rangle \}$ [Eq. \eqref{eq:LihntFrontbasisSet}], the matrix of the modified mass-squared operator for the $N\pi$ system [Eq. \eqref{eq:HwithCMregulated}] can be constructed. By solving the eigenequation (via numerical matrix diagonalization)
\begin{align}
H | \Psi _i \rangle =& M^2 _i | \Psi _i \rangle \ ,
\end{align}
we obtain the eigenmass $M_i$ and the corresponding eigenvector  
\begin{align}
| \Psi _i \rangle  \equiv &\sum _{\xi}   C_i ( \xi )\ | \xi \rangle \ \label{eq:BasisExpansionXX} ,
\end{align}
with $C _i ( \xi ) = \langle \xi | \Psi _i \rangle $ being the LF amplitude corresponding to the basis state $ | \xi \rangle $. The summation is taken over the LF basis set $\{ | \xi \rangle \}$. The LFWF \footnote{In principle, the application of the Fock-sector truncation requires the renormalization of the LFWF (see, e.g., Ref. \cite{Zhao:2014xaa}). We defer this study to a future effort.} is made up by the LF amplitudes $\{ \langle \xi | \Psi _i \rangle \}$. For computational efficiency, we limit the summation in Eq. \eqref{eq:BasisExpansionXX} to basis states of a specified symmetry as discussed above in Sec. \ref{sec:SymmtryAndBasisSelection}.  Separate calculations are then performed to obtain solutions of each desired symmetry.  

\subsubsection{Probability density distribution of the pion's longitudinal momentum fraction}
\label{sec:probabilityDensityDistributionObservables}
The probability to find a constituent pion of the longitudinal momentum fraction $x_{\pi}$ in our $N\pi$ model can be computed based on the LFWF, which is 
\begin{align}
\widetilde{f}_{\pi}(x_{\pi}) \equiv {\sum} ' C^{\ast} ( \xi  )  C ( \xi )  \ \label{eq:PDFequationxxx} ,  
\end{align} 
where it is understood that $x_{N} = 1 - x_{\pi}$ according to the conservation of the longitudinal momentum. The primed sum in Eq. \eqref{eq:PDFequationxxx} denotes that 1) the sum is over all the quantum numbers except $x_{\pi}$; 2) the sum includes only states with a pion; and 3) the sum is performed for the amplitudes corresponding to a selected mass eigenstate so that the index $i$ is suppressed. 

Correspondingly, we can also define the probability to find a constituent proton of the longitudinal momentum fraction $x_{N}$ as $\widetilde{f}_{N}(x_{N})$ for $x_N \neq 1$. \footnote{Recall that the zero mode of the pion field is excluded and we have $\widetilde{f}_{\pi}(x_{\pi}=0)=0$ throughout this work.} The following identity holds
\begin{align}
\widetilde{f}_{\pi}(x_{\pi}) = \widetilde{f}_{N}(x_{N}) \ \label{eq:PDFequationPrime}.
\end{align}

In this work, we rescale $\widetilde{f}_{\pi}(x_{\pi})$ by $1/K_{\rm max}$, which is the resolution in the longitudinal direction. The probability density distribution of the pion's longitudinal momentum fraction is hence defined as
\begin{align}
f_{\pi} (x_{\pi}) \equiv {K_{\rm max}} \cdot \widetilde{f}_{\pi}(x_{\pi}) \label{eq:PDFequation} \ .
\end{align}

According to $f_{\pi} (x_{\pi})$, we compute various related moments. The zero$^{\rm th}$ moment is 
\begin{align}
I^{(0)}_{\pi} \equiv \int _0^1 f_{\pi} (x_{\pi}) d x_{\pi}  \ \label{eq:zeroM} .
\end{align}
$I^{(0)}_{\pi}$ denotes the total probability to find the physical proton as a composite system of the bare nucleon and pion. On the other hand, the probability to find the bare nucleon is
\begin{align}
Z_2 \equiv \sum |\langle N_{\rm bare} | p \rangle _{\rm phys}|^2 =1- I^{(0)}_{\pi} \ \label{eq:Z2}.
\end{align}

The first moment is
\begin{align}
I^{(1)}_{\pi} \equiv & \int _0^1 x_{\pi}f_{\pi} (x_{\pi}) d x_{\pi}\ \label{eq:firstM} ,
\end{align}
which presents the average longitudinal momentum carried by the pion.
The second-moment with respect to the longitudinal momentum carried by the pion is
\begin{align}
I^{(2)}_{\pi} \equiv \int _0^1 x^2_{\pi}f_{\pi} (x_{\pi}) d x_{\pi}\ \label{eq:secondM} ,
\end{align}
which is related to the fluctuation of the pion's longitudinal momentum fraction in our model.

\subsubsection{Dirac form factor}
\label{sec:DiracFF}
In the LF coordinates, the Dirac form factor can be computed as \cite{Brodsky:2000ii}
\begin{align}
F_1 (Q^2) =& \frac{1}{2P^+} \langle P' , \uparrow | {J^+ (0)} | P, \uparrow \rangle \ ,
\end{align}
where the upward arrows denote the initial and final states with the projections of the total angular momenta being $+\frac{1}{2}$. $P$ and $P'$ are the momenta of the initial and finial states, respectively. $q= P' - P$ is the momentum carried by the probing virtual photon. Adopting the Drell-Yan frame, we have 
\begin{align}
q=\big( q^+, q^-, q^{\perp} \big) = \big( 0, - \frac{q^2}{P^+} , q^{\perp} \big) \ , \\
P=\big( P^+, P^-, P^{\perp} \big) = \big( P^+ , \frac{M^2}{P^+} , 0 \big) \ , \\
q^2 = -2 P \cdot q = -\big( q^{\perp} \big)^2 \ \equiv - Q^2 \ \label{eq:squaredTransMomentumTransfer},
\end{align}
where $Q^2$ is referred to as the squared transverse momentum transfer in the following. In principle, the Dirac form factor (or, more generally, observables) should be frame independent due to the Lorentz invariance. In practice, however, the Lorentz symmetry is  broken by the Fock-sector truncation in our model \cite{Carbonell:1998rj,Li:2017uug,Li:2019kpr}.\footnote{In this work, higher Fock sectors, such as $|N\pi \pi \rangle$, are omitted in Eq. \eqref{eq:fockSectorTruncation}. As the higher Fock sectors are systematically included in our model, the Lorentz symmetry can be dynamically restored, through which we anticipate that the Dirac form factor gradually becomes frame independent.} The frame dependence of the Dirac form factor could hence serve as a measure of the Lorentz symmetry violation, which will be the topic of a future work.

In our current model, the Fock-sector expansion for the physical proton can be schematically written as
\begin{align}
|p _{\rm phys} \rangle =& a_p |p \rangle + a_{p \pi ^0} | p \pi ^0 \rangle + a_{n \pi ^+} | n \pi ^+ \rangle \ \label{eq:fockSectorTruncationProton} \ , 
\end{align}
where $a_p$, $a_{p \pi ^0}$ and $a_{n \pi ^+}$ schematically represent the amplitudes since each term on the right hand side of Eq. \eqref{eq:fockSectorTruncationProton} represents a sum over the basis states with corresponding individual amplitudes. Hence, there are three different classes of contributions to the Dirac form factor of the physical proton: (1) the virtual photon couples to the current of the bare proton $|p \rangle$, which results in $F_{1,f} ^{p} (Q^2)$; (2) the virtual photon couples to the current of the bare proton when dressed by charge-neutral $\pi ^0$, which results in $F_{1,f} ^{p \pi ^0} (Q^2)$; and (3) the virtual photon couples to the current of $\pi ^+$, which results in $F_{1,b} ^{n \pi ^+} (Q^2) $. The Dirac form factor for the physical proton is hence
\begin{align}
F_{1} (Q^2) =& F_{1,f} ^{p} (Q^2) + F_{1,f} ^{p \pi ^0} (Q^2) + F_{1,b} ^{n \pi ^+} (Q^2) \ \label{eq:formfactorExtended},
\end{align}
where the subscripts $f$ and $b$ denote the contributions to $F_1(Q^2)$ from the fermionic current and the bosonic current, respectively. The detailed expression of $F_1(Q^2)$ is shown in Appendix \ref{sec:DiracFFformula}.

\subsubsection{Proton's r.m.s. charge radius}
The proton's r.m.s. charge radius $\sqrt{\langle r^2_{p,E} \rangle}$ can be calculated from the following expression \cite{Miller:2018ybm}
\begin{align}
\langle r^2_{p,E} \rangle = & -6 \frac{d G_E(Q^2)}{dQ^2} \Bigg|_{Q^2 \rightarrow 0} \ \label{eq:SlopeRMS} \ ,
\end{align}
where $G_E(Q^2)$ denotes the Sachs electric form factor 
\begin{align}
G_E(Q^2) =& F_1 (Q^2) - \frac{Q^2}{4M_N^2} F_2 (Q^2)\ \label{eq:SachsFF}.
\end{align}

By modification, we relate $\sqrt{\langle r^2_{p,E} \rangle}$ to the  slope of the proton Dirac form factor $F_1(Q^2)$ at vanishing $Q^2$ as
\begin{align}
{  \frac{d F_1(Q^2) }{dQ^2}  \Bigg|_{Q^2 \rightarrow 0} } = -\frac{1}{6} \langle r^2_{p,E} \rangle + \frac{1}{4M_N^2} { F_2 (0) } \ \label{eq:slopeofF1}.
\end{align}
Note that this slope is negative for the proton. In this work, we take $F_2 (0)=1.7928$ (in units of nuclear magneton $\mu _N$) \cite{Perdrisat:2006hj} for the proton when extracting the r.m.s. charge radius according to Eq. \eqref{eq:slopeofF1}. Refining this approach will be a future research effort.

\section{Results and discussions}
\label{sec:ResultsDiscussions}
In this work, we adopt the Fock-sector dependent renormalization (FSDR) \cite{Hiller:1998cv,Karmanov:2008br,Karmanov:2010ih,Karmanov:2012aj} scheme. We numerically diagonalize the matrix of the modified mass-squared operator $H$ [Eq. \eqref{eq:HwithCMregulated}] using an iterative process in which the bare nucleon mass is tuned in the matrix elements within the single-nucleon sector. This process continues until the square-root of the eigenvalue of the ground state (identified as a physical proton) matches the mass of the physical proton (taken as 938.272 MeV in this work). 

According to the FSDR scheme, the mass counterterm is introduced only to the single-nucleon sector. We expect the mass counterterm to compensate for the mass correction due to the radiative processes: the quantum fluctuation from the single-nucleon sector to the $N\pi$ sector and back again. On the other hand, the nucleon mass in the $N\pi$ sector remains as the physical value until a future effort would renormalize it with the inclusion of a higher Fock sector. In the FSDR procedure, we fix the pion mass as 137 MeV.

\subsection{Mass spectrum of the $N\pi$ system}
\begin{figure}[H]
\centering
\includegraphics[width=15cm]{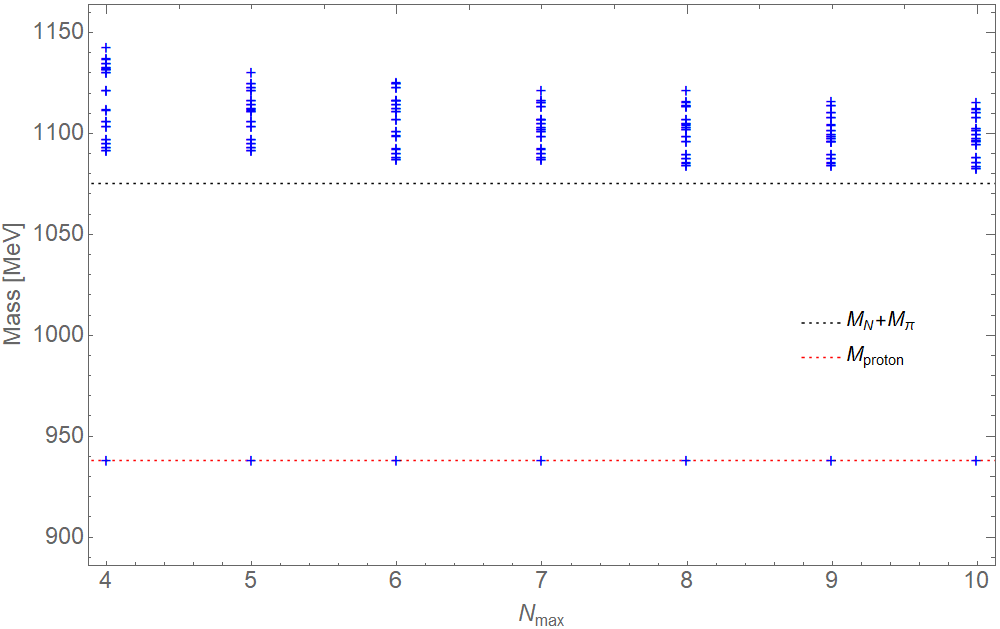}
\caption{Model space dependence of the mass spectrum of the $N\pi$ system computed via the BLFQ approach. The masses of the lowest 30 eigenstates are plotted as functions of $N_{\rm max}$. The basis strength is fixed as $b=250$ MeV for the purpose of demonstration. $K_{\rm max}$ is set to be 97/2 for good convergence. The dashed black line (at 1075.272 MeV) shows the continuum threshold of the $N\pi$ system. The ground state (bound) is identified as the physical proton, of which the mass is renormalized to $M_{\rm proton}=938.272$ MeV in the FSDR scheme. The ground states for all choices of  model spaces (labeled by $N_{\rm max}$) are joined by the dashed red line to guide the eye.
}
\label{fig:massSpectrum}
\end{figure}

As an illustration of how we solve the proton as the relativistic bound state of the $N\pi$ system, we present in Fig. \ref{fig:massSpectrum} the mass spectrum of the lowest 30 states of the $N\pi$ system as a function of the model space (scaled by $N_{\rm max}$, $b$ and $K_{\rm max}$). We set $K_{\rm max}=97/2$ (for good convergence) and $b=250$ MeV (for the simple purpose of demonstration). The Lagrangian multiplier in Eq. \eqref{eq:HwithCMregulated} is set to be $\Lambda =300$ MeV in this demonstration, such that no state with CM excitation is present in Fig. \ref{fig:massSpectrum}: according to the Lipkin-Lawson method [Eq. \eqref{eq:HwithCMregulated}], the eigenenergy of the lowest state with the CM excitation is 1238.272 MeV.

For each $N_{\rm max}$, we fit the ground state eigenvalue to be 938.272 MeV, which we identify as the physical proton. The corresponding wave function is identified as the proton LFWF, which is boost invariant. The other states lie above the threshold of the continuum (dashed black line in Fig. \ref{fig:massSpectrum}), which is the sum of the physical pion and nucleon masses adopted in this work (i.e., 1075.272 MeV); they represent the $N\pi$ scattering states. As $N_{\rm max}$ increases (i.e., as more basis states are added), a better representation of the scattering states of the $N\pi$ system will emerge. This can be inferred from the increasing level density of the scattering states as $N_{\rm max}$ increases. We defer detailed investigation of the continuum states, such as the convergence of sum rules (c.f., Ref. \cite{Dinur:2014kha}), to a later effort.

\subsection{Choices of the model space parameters for the proton's LFWFs}
\label{sec:choice}
To compute the observables such as the probability density distribution of the pion's longitudinal momentum fraction $f_{\pi}(x_{\pi})$ and the proton's Dirac form factor $F_1(Q^2)$, we determine the model space parameters ($N_{\rm max}$, $b$, $K_{\rm max}$) in our calculations as following. For each $N_{\rm max}$, 1) we determine the mass counterterm by fitting proton mass (938.272 MeV) via the FSDR procedures; 2) we choose $b$ by fitting the proton's r.m.s. charge radius $\sqrt{\langle r^2_{p,E} \rangle}$ (adopted to be either 0.879 fm \cite{Arrington:2015ria} or 0.840 fm \cite{Bezginov:2019mdi} in view of the proton radius puzzle); 3) we select a sufficiently large $K_{\rm max}$ for good convergence; 4) finally, we restrict the 2DHO basis parameters such that the IR/UV cutoffs of the basis space (Eqs. \eqref{eq:UVcut} and \eqref{eq:IRcut}) are consistent with a scale assumed to be reasonable for the chiral effective field theory we investigate. 

\begin{table}[H]
\caption{Basis parameters employed to obtain the proton's LFWFs. We take the proton r.m.s. charge radius to be $\sqrt{\langle r^2_{p,E} \rangle} = 0.879$ fm \cite{Arrington:2015ria} in this case. The corresponding UV and IR cutoffs of the 2DHO basis as defined in Eqs. \eqref{eq:UVcut} and \eqref{eq:IRcut} are also presented. $K_{\rm max}$ is set to be $97/2$ for good convergence.}
\centering
\begin{tabu}{c c c c c c } 
 \hline\hline
 $N_{\rm max}$ & $K_{\rm max}$ & $b$ [MeV] & IR [MeV] & UV [MeV]  \\ \hline
 6 & $\frac{97}{2}$ & 157.169 &  45.371 & 544.450   \\ 
 7 & $\frac{97}{2}$ & 222.582 &  59.488 & 832.827   \\
 8 & $\frac{97}{2}$ & 228.871 &  57.218 & 915.483   \\
 \hline\hline
\end{tabu}
\label{table:parameterSettting}
\end{table}

\begin{table}[H]
\caption{Basis parameters employed to obtain the proton's LFWFs. We take $\sqrt{\langle r^2_{p,E} \rangle} = 0.840$ fm. The corresponding UV and IR cutoffs of the 2DHO basis as defined in Eqs. \eqref{eq:UVcut} and \eqref{eq:IRcut} are also presented. $K_{\rm max}$ is set to be $97/2$ for good convergence.}
\centering
\begin{tabu}{c c c c c c } 
 \hline\hline
 $N_{\rm max}$ & $K_{\rm max}$ & $b$ [MeV] & IR [MeV] & UV [MeV]  \\ \hline
 6 & $\frac{97}{2}$ & 201.858 &  58.271 & 699.257   \\ 
 7 & $\frac{97}{2}$ & 259.175 &  69.267 & 969.743   \\
 8 & $\frac{97}{2}$ & 262.096 &  65.524 & 1048.385   \\
 \hline\hline
\end{tabu}
\label{table:parameterSettting_SmallRMSfit}
\end{table}

In Tables \ref{table:parameterSettting} and \ref{table:parameterSettting_SmallRMSfit}, we present the basis parameter settings, along with the corresponding IR and UV cutoffs of the basis space. The parameter settings in Table \ref{table:parameterSettting} (Table \ref{table:parameterSettting_SmallRMSfit}) are obtained with the choice of $\sqrt{\langle r^2_{p,E} \rangle} = 0.879$ (0.840) fm. It is worth noting that, for obtaining smaller $\sqrt{\langle r^2_{p,E} \rangle}$, larger basis strength $b$ is required: for example, we have $b=157.169$ MeV for the choice of $N_{\rm max}=6$ in Table \ref{table:parameterSettting}, while $b$ is 201.858 MeV for the same $N_{\rm max}$ in Table \ref{table:parameterSettting_SmallRMSfit}. This can be understood from the fact that $1/b$ scales the typical length scale of the 2DHO basis, which is directly related to the characteristic size of the system. 

We attempted calculations with $N_{\rm max} \leq 5$ but were unsuccessful in obtaining both the physical mass and the proton r.m.s. charge radius. We also exclude the choices with $N_{\rm max} \geq 9$, since the UV cutoffs of such basis spaces are well above 1 GeV, which seems to be a reasonable UV limit for our chiral effective field theory. 

\subsection{Probability density distribution of pion's longitudinal momentum fraction}
\label{sec:NewResults}

\begin{figure}[H]
\centering
\includegraphics[width=10cm]{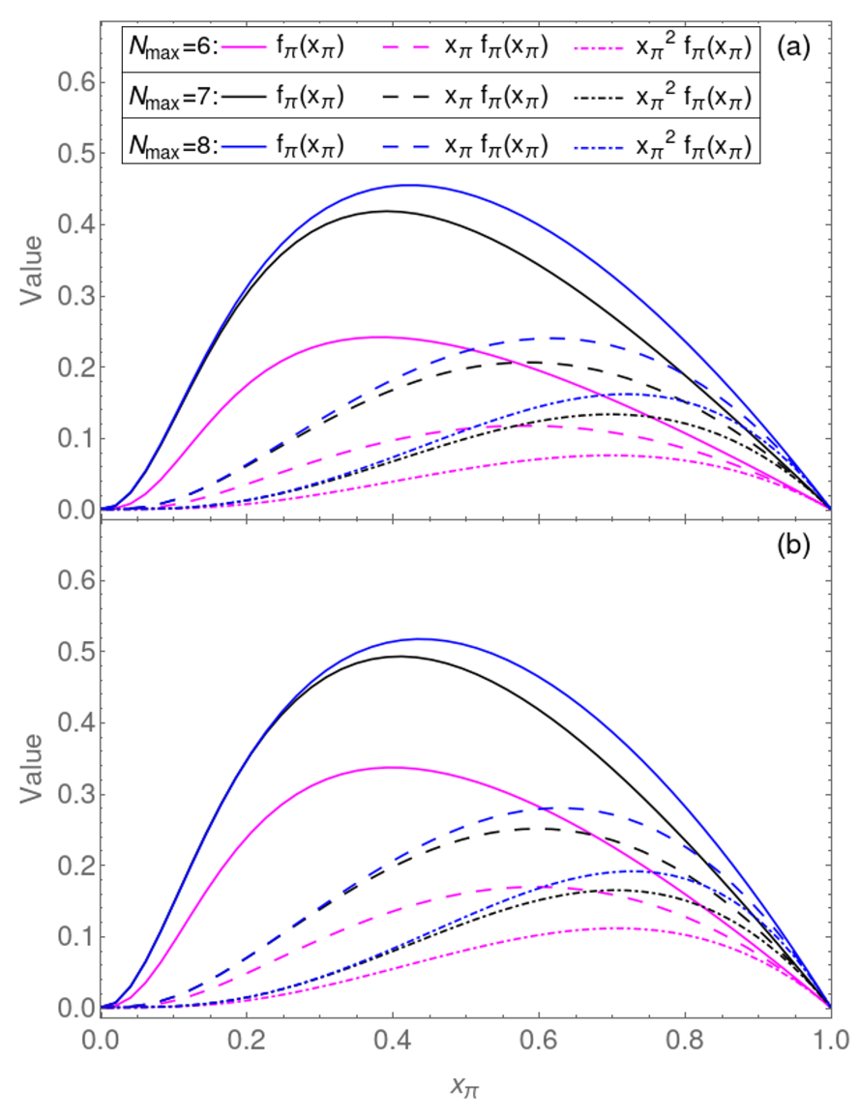}
\caption{(Color online) The quantities $f_{\pi}(x_{\pi})$ (solid lines),  $x_{\pi}f_{\pi}(x_{\pi})$ (dashed lines), and $x^2_{\pi}f_{\pi}(x_{\pi})$ (dot-dashed lines) as functions of the model space and the pion's longitudinal momentum fraction $x_{\pi}$. $N_{\rm max}$ labels the model space applied to compute the proton's LFWF: the magenta, black and blue lines denote the model spaces with $N_{\rm max}=6$, 7 and 8, respectively. For panel (a), the LFWFs are computed based on the parameter settings in Table \ref{table:parameterSettting}, where $b$ is fitted according to the choice of $\sqrt{\langle r^2_{p,E} \rangle}=$ 0.879 fm. For panel (b), the LFWFs are computed based on the parameter settings in Table \ref{table:parameterSettting_SmallRMSfit}, where $b$ is fitted according to the choice of $\sqrt{\langle r^2_{p,E} \rangle}=$ 0.840 fm.  
}
\label{fig:pionLongiFracBoth}
\end{figure}

In Fig. \ref{fig:pionLongiFracBoth}, we present the quantities $f_{\pi}(x_{\pi})$, $x_{\pi} f_{\pi}(x_{\pi})$ and $x^2_{\pi} f_{\pi}(x_{\pi})$ as functions of the model space (labeled by $N_{\rm max}$) and the pion's longitudinal momentum fraction $x_{\pi}$. The detailed definitions of these quantities are in Sec. \ref{sec:probabilityDensityDistributionObservables}. For Fig. \ref{fig:pionLongiFracBoth}(a) (Fig. \ref{fig:pionLongiFracBoth}(b)), the LFWFs are computed based on the parameter settings in Table \ref{table:parameterSettting} (Table \ref{table:parameterSettting_SmallRMSfit}). Recall that the model space parameters in Table \ref{table:parameterSettting_SmallRMSfit} are obtained by fitting to a smaller $\sqrt{\langle r^2_{p,E} \rangle}$, where the spatial extension of the proton is more restricted (correspondingly, a larger UV cutoff is needed).

We find that $f_{\pi}(x_{\pi})$ peaks at approximately $x_{\pi} \approx 0.40 \ (0.45)$ for the calculation with $N_{\rm max}=6 \ (8)$ in both Fig. \ref{fig:pionLongiFracBoth}(a) and Fig. \ref{fig:pionLongiFracBoth}(b). These peak positions indicate that, at the low momentum scale in this work (discussed below), the longitudinal momentum fraction carried by the constituent pion is more relativistic than na\"ive expectations ($x_{\pi} \approx 0.2$). 

In addition, we note that 1) the peak value increases with the model space dimensions (scaled by $N_{\rm max}$) for the calculations with a given choice of $\sqrt{\langle r^2_{p,E} \rangle}$; and 2) the peak value of $f_{\pi}(x_{\pi})$ are sensitive to the choice of $\sqrt{\langle r^2_{p,E} \rangle}$: those peak values in Fig. \ref{fig:pionLongiFracBoth}(b) are larger than their counterparts in Fig. \ref{fig:pionLongiFracBoth}(a). For example, at $N_{\rm max}=7$, the computed peak value of $f_{\pi}(x_{\pi})$ with the choice of $\sqrt{\langle r^2_{p,E} \rangle}=0.879$ fm is about 0.42, while the peak value computed with the choice of $\sqrt{\langle r^2_{p,E} \rangle}=0.840$ fm is about 0.50. These features of $f_{\pi}(x_{\pi})$ can be partly understood from the UV cutoff of the 2DHO basis: the peak value increases as the UV cutoff increases (Tables \ref{table:parameterSettting} and \ref{table:parameterSettting_SmallRMSfit}). 

In Figs. \ref{fig:pionLongiFracBoth}(a) and \ref{fig:pionLongiFracBoth}(b), we also present the results of $x_{\pi} f_{\pi}(x_{\pi})$ and $x^2_{\pi} f_{\pi}(x_{\pi})$. As expected from the properties of $f_{\pi}(x_{\pi})$ discussed above, the peak positions are about the same for the model spaces with same $N_{\rm max}$, though respective peak values are different.

We remark that the current $f_{\pi}(x_{\pi})$ results (and other related quantities), are calculated with the point-particle assumption, i.e., the constituent particles are treated as point-like, structureless particles. While including higher-order terms in the Lagrangian is necessary to improve the current calculations, it will also be important to account for the quark and gluon distributions within our constituent fields in order to enhance agreement with experiment at intermediate momentum transfers (e.g., $Q^2 \approx 4.0 \  {\rm GeV}^2$ or higher).

\begin{table}[H]
\caption{The zero$^{\rm th}$-, first- and second-moments of the pion's longitudinal momentum fraction. The model space parameters to calculate the LFWFs are shown in Table \ref{table:parameterSettting}, where the proton's $\sqrt{\langle r^2_{p,E} \rangle}$ is fitted to 0.879 fm. 
}
\centering
\begin{tabu}{c c c c} 
 \hline\hline
 $N_{\rm max}$  & $I^{(0)}_{\pi}$ & $I^{(1)}_{\pi}$ & $I^{(2)}_{\pi}$  \\ \hline
 6 & 0.144  & 0.068 & 0.039   \\ 
 7 & 0.254  & 0.120 & 0.068   \\
 8 & 0.285  & 0.138 & 0.080    \\
 \hline\hline
\end{tabu}
\label{table:moments}
\end{table}

\begin{table}[H]
\caption{The zero$^{\rm th}$-, first- and second-moments of the pion's longitudinal momentum fraction. The model space parameters to calculate the LFWFs are shown in Table \ref{table:parameterSettting_SmallRMSfit}, where the proton's $\sqrt{\langle r^2_{p,E} \rangle}$ is fitted to 0.840 fm.
}
\centering
\begin{tabu}{c c c c} 
 \hline\hline
 $N_{\rm max}$  & $I^{(0)}_{\pi}$ & $I^{(1)}_{\pi}$ & $I^{(2)}_{\pi}$  \\ \hline
 6 & 0.206  & 0.098 & 0.056   \\ 
 7 & 0.303  & 0.145 & 0.083   \\
 8 & 0.328  & 0.161 & 0.094    \\
 \hline\hline
\end{tabu}
\label{table:moments_smallerRMSfit}
\end{table}

Utilizing these distribution functions ($f_{\pi}(x_{\pi})$, $x_{\pi} f_{\pi}(x_{\pi})$ and $x^2_{\pi} f_{\pi}(x_{\pi})$), we calculate the zero$^{\rm th}$-, first- and second-moments of the pion's longitudinal momentum fraction (shown in Tables 
\ref{table:moments} and \ref{table:moments_smallerRMSfit}). In principle, individual moments can be viewed as the integrated area between the profile of the corresponding distribution function and the $x$-axis in Figs. \ref{fig:pionLongiFracBoth}(a) and \ref{fig:pionLongiFracBoth}(b). Physically, as explained in Sec. \ref{sec:probabilityDensityDistributionObservables}, $I^{(0)}_{\pi}$ and $I^{(1)}_{\pi}$ represent the probability of the $|N\pi \rangle$ sector and the average longitudinal momentum fraction carried by the constituent pion, respectively. $I^{(2)}_{\pi}$ is related to the quantum fluctuation of the pion's longitudinal momentum fraction. 

We note that the moments shown in Tables \ref{table:moments} and \ref{table:moments_smallerRMSfit} present the same patterns as those of the peak value of the $f_{\pi}(x_{\pi})$. For example, we find that 1) for the calculations with a given choice of $\sqrt{\langle r^2_{p,E} \rangle}$, the resulting $I^{(0)}_{\pi}$ increases with the model space dimensions (scaled by $N_{\rm max}$); 2) for a given $N_{\rm max}$, $I^{(0)}_{\pi}$ increases for the calculation based on the LFWF fitted to a smaller $\sqrt{\langle r^2_{p,E} \rangle}$. The same is true for $I^{(1)}_{\pi}$ and $I^{(2)}_{\pi}$. These features of the moments result directly from the properties of $f_{\pi}(x_{\pi})$ and could also be interpreted in terms of the sensitivity to the UV cutoff of the model space representation of the LFWFs.

\subsection{Probabilities of Fock sectors}
Based on the proton's LFWFs, we compute the probability of each Fock sector according to Eq. \eqref{eq:fockSectorTruncationProton}, as shown in Tables \ref{table:FockSectorProbabilitiesNoAuxiliaryFunc0879} and \ref{table:FockSectorProbabilitiesNoAuxiliaryFunc0840}. For each model space (Tables \ref{table:parameterSettting} and \ref{table:parameterSettting_SmallRMSfit}), we find that the probability of the bare proton sector ($|a_p|^2$) dominates, while the probability of the $|n\pi ^+ \rangle $ sector ($|a_{n\pi ^+}|^2$) is twice of that of the $|p\pi ^0 \rangle $ sector ($|a_{p\pi ^0}|^2$) due to the isospin symmetry \cite{Thomas:2007bc}. We also checked the normalization of the Fock sector expansion, i.e., $|a_p|^2+|a_{n\pi ^+}|^2+|a_{p\pi ^0}|^2=1$. (Note that there are nominal offsets from this sum in the data in Tables \ref{table:FockSectorProbabilitiesNoAuxiliaryFunc0879} and \ref{table:FockSectorProbabilitiesNoAuxiliaryFunc0840} due to round-off.) We find that the Fock-sector probabilities are sensitive to the choice of the proton's $\sqrt{\langle r^2_{p,E} \rangle}$. For the choice of smaller $\sqrt{\langle r^2_{p,E} \rangle}$ (i.e., 0.840 fm), the resulting $|a_p|^2$ becomes less dominant (while $|a_{n\pi ^+}|^2$ and $|a_{p\pi ^0}|^2$ becomes more important). 

With the choice of the proton's $\sqrt{\langle r^2_{p,E} \rangle}$ to be either 0.879 fm or 0.840 fm, we find that the total probability of a physical proton consisting of a bare nucleon and a single pion is within the range of $[0.144,0.285]$ (Table \ref{table:moments}) or the range $[0.206,0.328]$ (Table \ref{table:moments_smallerRMSfit}). Correspondingly, the probability of the $| n\pi^+  \rangle$ sector is within the range of $[0.096,0.190]$ (Table \ref{table:FockSectorProbabilitiesNoAuxiliaryFunc0879}) or the range of $[0.137,0.218]$ (Table \ref{table:FockSectorProbabilitiesNoAuxiliaryFunc0840}), while the probability of the $| p\pi ^0 \rangle $ sector is within the range of $[0.048,0.095]$ (Table \ref{table:FockSectorProbabilitiesNoAuxiliaryFunc0879}) or the range of $[0.069,0.109]$ (Table \ref{table:FockSectorProbabilitiesNoAuxiliaryFunc0840}). These results compare well with the studies of the pion cloud effects by the cloudy bag model \cite{Thomas:2007bc,Alberg:2012wr}. However, it is also worth noting that the correct size of the pion cloud effects would not be sufficient to obtain the electromagnetic form factors, as will be discussed below.

\begin{table}[H]
\caption{Fock-sector probabilities [Eq. \eqref{eq:fockSectorTruncationProton}] computed from the proton's LFWFs. The model space parameters to calculate the LFWFs are shown in Table \ref{table:parameterSettting}, where the proton's $\sqrt{\langle r^2_{p,E} \rangle}$ is fitted to 0.879 fm.
}
\centering
\begin{tabu}{c c c c} 
 \hline \hline
 $N_{\rm max}$ & $|a_p|^2$ & $|a_{n\pi ^+}|^2$  & $|a_{p\pi ^0}|^2$ \\ \hline
 6  & 0.856 & 0.096 & 0.048 \\ 
 7  & 0.746 & 0.169 & 0.085 \\
 8  & 0.715 & 0.190 & 0.095 \\
 \hline \hline
\end{tabu}
\label{table:FockSectorProbabilitiesNoAuxiliaryFunc0879}
\end{table}

\begin{table}[H]
\caption{Fock-sector probabilities [Eq. \eqref{eq:fockSectorTruncationProton}] computed from the proton's LFWFs. The model space parameters to calculate the LFWFs are shown in Table \ref{table:parameterSettting_SmallRMSfit}, where the proton's $\sqrt{\langle r^2_{p,E} \rangle}$ is fitted to 0.840 fm.}
\centering
\begin{tabu}{c c c c} 
 \hline \hline
 $N_{\rm max}$ & $|a_p|^2$ & $|a_{n\pi ^+}|^2$  & $|a_{p\pi ^0}|^2$ \\ \hline
 6  & 0.794 & 0.137 & 0.069 \\ 
 7  & 0.697 & 0.202 & 0.101 \\
 8  & 0.672 & 0.218 & 0.109 \\
 \hline \hline
\end{tabu}
\label{table:FockSectorProbabilitiesNoAuxiliaryFunc0840}
\end{table}

\subsection{Proton's Dirac form factor}
\label{sec:DiracFFresults}

\begin{figure}[H]
\centering
\includegraphics[width=14cm]{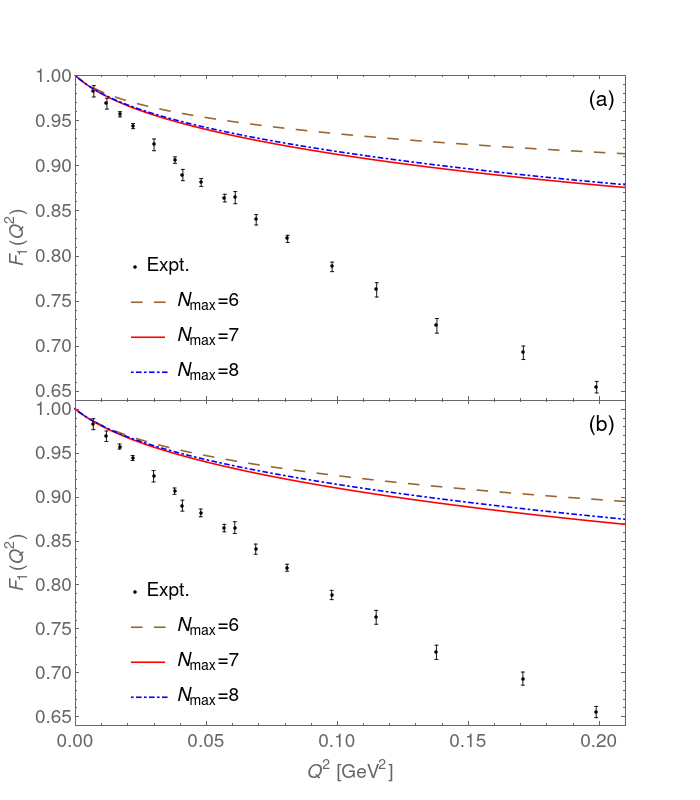}
\caption{(Color online) The computed Dirac form factor $F_1(Q^2)$ of the proton as a function of the squared transverse momentum transfer $Q^2$ and of the model space. $N_{\rm max}$ labels the model space applied to compute the proton's LFWF: the brown dashed line, the red solid line and the blue dot-dashed line denote results based on the model spaces with $N_{\rm max}=6$, 7 and 8, respectively. The experimental results (black) \cite{Arrington:2007ux}, along with the corresponding error bars, are also presented for comparison.  
For panel (a), the LFWFs are computed based on the parameter settings in Table \ref{table:parameterSettting}, where $b$ is fitted according to the choice of $\sqrt{\langle r^2_{p,E} \rangle}=$ 0.879 fm. For panel (b), the LFWFs are computed based on the parameter settings in Table \ref{table:parameterSettting_SmallRMSfit}, where $b$ is fitted according to the choice of $\sqrt{\langle r^2_{p,E} \rangle}=$ 0.840 fm. 
}
\label{fig:FormFactorFull}
\end{figure}

We apply the boost invariant LFWFs (computed based on the parameter settings in Table \ref{table:parameterSettting}) to calculate the proton's Dirac form factor. In Fig. \ref{fig:FormFactorFull}, we present the proton's Dirac form factor $F_1(Q^2)$ [Eq. \eqref{eq:formfactorExtended}] as a function of the squared transverse momentum transfer $Q^2$ [Eq. \eqref{eq:squaredTransMomentumTransfer}] and of the model space (labeled by $N_{\rm max}$) with parameters listed in Tables \ref{table:parameterSettting} and \ref{table:parameterSettting_SmallRMSfit}. 

In computing the proton's LFWFs, we tune $b$ for each choice of $N_{\rm max}$ such that the LFWFs produce the adopted $\sqrt{\langle r^2_{p,E} \rangle}$ values (along with the physical proton's mass). This can be seen from the agreement between the slopes of $F_1(Q^2)$ results and the experimental result in the vicinity of the origin.

The results of $F_1(Q^2)$ with the same $N_{\rm max}$ in Fig. \ref{fig:FormFactorFull}(a) and Fig. \ref{fig:FormFactorFull}(b) are similar, except for the different rates of decrease with increasing $Q^2$. 
In the limit of vanishing $Q^2$, we find $F_1(0)=1$ for all of the results, which indicates the conservation of charge. This also shows a proper normalization of the proton's LFWFs: the normalization is precise to at least eight significant figures in the current work. In the limit of $Q^2 \rightarrow \infty$, the $F_1(\infty)$ is equal to the probability of the bare proton sector $|a_p|^2$ (Eq. \eqref{eq:fockSectorTruncationProton}, Tables \ref{table:FockSectorProbabilitiesNoAuxiliaryFunc0879} and \ref{table:FockSectorProbabilitiesNoAuxiliaryFunc0840}).

For the present application, we expect our calculation to be valid only for small momentum scale $Q^2 < m^2_{\pi}$ (where the point-particle assumption should be reasonable). Indeed, we find from Fig. \ref{fig:FormFactorFull} that the computed $F_1(Q^2)$ deviates significantly from the experimental results as $Q^2$ increases. While the restricted Fock-sector truncation [Eq. \eqref{eq:fockSectorTruncationProton}] may be one reason, the major drawback is the simplicity of our current chiral model. As reported in the review \cite{Perdrisat:2006hj}, inclusion of the vector mesons and also the $\Delta$-resonance state of pion and nucleon can increase the range of agreement up to about $Q^2=0.16\ {\rm GeV}^2$. To achieve the agreement for even higher $Q^2$, more fundamental degrees of freedom (quarks and gluons) need to be included (e.g., Ref. \cite{Miller:2002ig}).

\section{Conclusions and outlook}
\label{sec:ConclusionOutlook}
We apply, for the first time, the Basis Light-Front Quantization (BLFQ) approach \cite{Vary:2009gt} to study a chiral model for the nucleon-pion ($N\pi$) system via a non-perturbative, Hamiltonian approach. We consider a model problem, where a physical proton is treated as the relativistic bound state of the $N\pi$ system.

Starting from the Lagrangian density for the chiral model of the $N\pi$ system (c.f., \cite{Miller:1997cr,Miller:2000kv}), we proceed with the Legendre transformation to obtain the corresponding light-front (LF) Hamiltonian density. In this work, we keep only the Fock sectors $|N \rangle$ and $|N\pi \rangle$. Correspondingly, we restrict the interaction terms in the LF Hamiltonian density and keep only the terms that correspond to the single-pion emission and absorption processes. 

We then show our choice of the construction and truncation schemes of the LF basis. In particular, we employ the discretized plane wave basis in the longitudinal direction and the two dimensional harmonic oscillator basis in the transverse direction. Besides the basis sets in momentum space, we also discuss our choice of the basis set in spin and isospin degrees of freedom. We prune our basis according to the symmetries of the Hamiltonian for our chosen system.

We construct the matrix of the mass-squared operator within the LF basis representation, where we regulate the center of mass excitation by the Lipkin-Lawson method \cite{Lipkin:1958zza,Gloeckner:1974sst}. Incorporating the Fock-sector-dependent renormalization (FDSR) \cite{Hiller:1998cv,Karmanov:2008br,Karmanov:2010ih,Karmanov:2012aj} scheme, we obtain the mass spectrum of the proton and the corresponding boost-invariant light-front wave function (LFWF) by solving the eigenvalue problem of the mass-squared operator. 

We first illustrate the mass spectrum of the $N\pi$ system in the BLFQ approach. The mass spectrum includes both the bound and scattering states. We present the lowest 30 states as a function of the model space, which is determined by the truncation parameters ($N_{\rm max}$ and $K_{\rm max}$), basis strength ($b$), and the current choice of Fock sectors. We find that the eigenvalue of the ground state produces the physical proton mass for each model space with proper choice of the mass counterterm. The remaining 29 states represent the scattering states of the $N\pi$ system. By increasing the model-space dimensionality, a better representation of the continuum could be obtained for the current $N\pi$ system.

We then show our solution of the selected observables, i.e., the probability density distribution of the pion's longitudinal momentum fraction $f_{\pi} (x_{\pi})$, the quantities relating to $f_{\pi} (x_{\pi})$, and the proton's Dirac form factor $F_1(Q^2)$. To this end, we compute the proton's LFWFs in a sequence of model spaces (determined by $N_{\rm max}$, $K_{\rm max}$ and $b$) where both the proton's mass and its r.m.s. charge radius are fitted to respective experimental values. We select to fit the proton's r.m.s. charge radius to be either 0.879 or 0.840 fm, in view of the proton radius puzzle (see e.g., Refs. \cite{Miller:2018ybm,Bezginov:2019mdi} and the references therein). We find that the longitudinal momentum fraction carried by the constituent pion is more relativistic that na\"ive expectations at the low momentum scale. In addition, the $f_{\pi} (x_{\pi})$ results are sensitive to the choice of the proton's r.m.s. charge radius. We will defer the efforts of improving the $f_{\pi} (x_{\pi})$ calculations to future works.

The same sets of the proton's LFWFs are applied to compute the proton's Dirac form factor $F_1(Q^2)$. We study the proton's Dirac form factor $F_1(Q^2)$ as a function of the squared transverse momentum transfer $Q^2$ and the model space. For all choices of model space, the results of $F_1(Q^2)$ agree well with the experimental results when the momentum scale is small $Q^2 < m^2_{\pi}$. As $Q^2$ increases, our results of $F_1(Q^2)$ deviate from the experimental results. We argue this is mainly because of the simplicity of the chiral model in this work: both species of the constituent particles (nucleons and pions) are assumed to be point-like particles. 

This work can lead to a number of pathways for further research. We attempt to connect the current chiral model to the modern chiral effective theory (see, e.g., \cite{Entem:2003ft,Epelbaum:2008ga,Machleidt:2011zz} and references therein). This work is currently ongoing. After this connection is accomplished, we plan to extend the current calculation to incorporate systematically the contributions from higher Fock sectors, where we will examine the basis-space dependence as well as the convergence of the Fock-sector expansion. We expect such investigations to be demanding in computing power. To address this difficulty, we plan to incorporate the technology of high performance computing (see Ref. \cite{Vary:2018hdv} and references therein). 

The current framework can also be straightforwardly extended to investigate more nucleonic observables of experimental interest, such as the generalized parton distribution, the transverse momentum distribution, and various types of form factors (especially, the nucleon axial form factors that are of high current interest for neutrino physics \cite{Bodek:2007vi,Bernard:2001rs,Formaggio:2013kya}). In addition, this framework can be extended to study more complicated nuclear systems, such as the deuteron, where the role of the relativistic dynamics is important but still unclear.

\section*{Acknowledgments}
We acknowledge valuable discussions with A. W. Thomas, M. Burkardt, P. Maris, K. Tuchin, T. Frederico and L. Geng. This work was supported by the U.S. Department of Energy (DOE) under grant No. DE-FG02-87ER40371. X. Zhao is supported by new faculty startup funding by the Institute of Modern Physics, Chinese Academy of Sciences and by Key Research Program of Frontier Sciences, CAS Grant No. ZDBS-LY-7020.

\begin{appendices}
\section{2DHO basis}
\label{sec:2DHObasis}
The generating operator for the 2DHO basis can be expressed as \cite{Wiecki:2014ola}
\begin{align}
P_+ ^{\Omega} =& \frac{(p^{\perp})^2}{2p^+} + \frac{1}{2} {\Omega}^2 p^+ (r^{\perp})^2 \ = \ \frac{1}{2} \Omega \Big[ \frac{(p^{\perp})^2}{xP^+ \Omega} + x P^+ \Omega (r^{\perp})^2 \Big] \ ,
\end{align}
where the oscillator energy $\Omega$ is related to the energy scale of the 2DHO basis set as
\begin{align}
b= \sqrt{P^+ \Omega} \ .
\end{align}
In the following, we refer to $b$ as the basis strength. 

For the convenience in evaluating integrals involving the 2DHO basis, one can further introduce the momentum fraction weighted variables \cite{Maris:2013qma} as
\begin{align}
q^{\perp} \equiv  \frac{p^{\perp}}{\sqrt{x}}  , \ s^{\perp} \equiv \sqrt{x} r^{\perp} \ \label{eq:MomentumFractionWeightedVariables},
\end{align} 
where the canonical commutator $[s_i^{\perp}, q_j^{\perp}]=i\delta _{ij}$ ($i,j=1,2$)  holds. The generating operator of the 2DHO basis in terms of the conjugate variables $(s^{\perp}, q^{\perp})$ can be rewritten as
\begin{align}
P_+ ^{\Omega} =& \frac{1}{2} \Omega \Big[ \Big(  \frac{q^{\perp}}{\sqrt{P^+ \Omega}} \Big)^2 + \Big( \sqrt{P^+ \Omega} s^{\perp} \Big)^2 \Big] \ .
\end{align}
In the momentum representation, the 2DHO wave function is
\begin{align}
\langle q^{\perp} | nm \rangle \ = \ \Psi _n^m (q^{\perp})\ = \ \frac{1}{b} \sqrt{\frac{4 \pi n!}{(n+|m|)!}} \rho ^{|m|} e^{-\frac{1}{2} \rho ^2} L_n ^{|m|} (\rho ^2)\ e ^{im \phi} \ \label{eq:2DHOinComplexMomentumRep},
\end{align}
where the transverse momentum in the complex representation is
\begin{align}
& q^{\perp} = b \rho e^{i \phi} \ , 
\end{align}
with $ \phi = \mathrm{arg}\ q^{\perp}, \ |q^{\perp}| = b \rho $. Correspondingly, we have $(q^{\perp})^{\ast} = b \rho e^{-i \phi}$. $n$, $m$ are the quantum numbers for the radial part and angular part of the wave function, respectively. They define the eigenenergy of the corresponding 2DHO wave function
\begin{align}
E_{nm} =& (2n+ |m|+1) \Omega \ .
\end{align}
$L_n ^{|m|} (\rho ^2)$ denotes the generalized Laguerre polynomial. 

The orthonormality relation of the 2DHO basis is
\begin{align}
\langle nm| n'm' \rangle \equiv  \int \frac{d ^2 q^{\perp}}{(2\pi)^2} \Psi _n^{m \ast} (q^{\perp}) \Psi _{n'}^{m'} (q^{\perp}) = \delta _{n,n'} \delta _{m,m'} \ .
\end{align}

\section{Contributions to the LF Hamiltonian}
\label{sec:ContributionToLFHamiltonian}
\subsection{Kinetic energy for the $N\pi$ system}
The contribution from a free nucleon to the LF Hamiltonian $P^-$ is
\begin{align}
P^-_{\rm KE_N} =& \sum _{p^+} \sum _{s = -\frac{1}{2}} ^{\frac{1}{2}} \sum _{t=-\frac{1}{2}} ^{\frac{1}{2}} \int \frac{d^2 p^{\perp}}{(2\pi)^2}  b^{\dagger} (p , s, t )\ b (p , s, t )  \frac{(p^{\perp})^2 +M_N^2}{p^+} \ .
\end{align}
Substituting Eq. \eqref{eq:bToBeta} to the above expression, we obtain the analytic expression of the contribution of a free nucleon to the mass squared operator $H_{\rm LC}$ [Eq. \eqref{eq:HLCoperator}] in terms of the LF basis:
\begin{align}
& P^+ P^-_{KE_N} \nonumber \\
 =&  \sum _{x_N} \sum _{s} \sum _{t} \sum _{n_1,m_1} \sum _{n_2,m_2}  \beta ^{\dagger} (x_N, n_1,m_1,s , t) \beta  (x_N, n_2,m_2,s , t) \ \delta _{m_1,m_2} \nonumber \\
& \times \Bigg\{ { b^2  \Big[ (2n_2 + |m_2|+1) \delta _{n_1,n_2} - \sqrt{n_1(n_1+|m_1|)} \delta _{n_1,n_2 +1} - \sqrt{n_2(n_2+|m_2|)} \delta _{n_2 ,n_1+1} \Big] + \frac{M_N^2}{x_N}  \delta _{n_1,n_2} } \Bigg\} \ \label{eq:NKEtoTheMassSquareOperator} ,
\end{align}
where we have applied the relation $x_N=\frac{p^+}{P^+}$ [Eq. \eqref{eq:singleParticleRelation2}] with $P^+$ being the total longitudinal momentum. Note that $x_N=1$ for the $ | N \rangle $ sector and $0<x_N<1$ for the $|N\pi \rangle $ sector. When evaluating the integral, we have also made use of the integral identity Eq. \eqref{eq:KEterm2DHOintegral}.

The contribution from a free pion to $P^-$ is
\begin{align}
P^-_{\rm KE_{\pi}} =&  \sum _{k^+} \sum _{\lambda =-1}^{1} \int \frac{d^2 k^{\perp}}{(2\pi)^2}  a^{\dagger} (k, \lambda) \ a (k, \lambda)  \frac{(k^{\perp})^2+M_{\pi}^2}{k^+} \ .
\end{align} 
Analogous to Eq. \eqref{eq:NKEtoTheMassSquareOperator}, we also obtain the expression of the contribution of a free pion to the mass squared operator $H_{\rm LC}$ [Eq. \eqref{eq:HLCoperator}]:
\begin{align}
& P^+ P^-_{KE_{\pi}} \nonumber \\
=&  \sum _{x_{\pi}} \sum _{\lambda} \sum _{n_1 m_1} \sum _{n_2 m_2} \alpha ^{\dagger} (x_{\pi}, n_1, m_1, \lambda) \alpha (x_{\pi}, n_2, m_2,\lambda) \ \delta _{m_1,m_2} \nonumber \\
& \times  \Bigg\{ { b^2 \Big[ (2n_2 +|m_2|+1) \delta_{n_1 n_2} - \sqrt{n_1 (n_1 +|m_1|)} \delta _{n_1, n_2 +1}  - \sqrt{n_2 (n_2 +|m_2|)} \delta _{n_2,n_1+1} \Big] + \frac{M_{\pi}^2}{{x_{\pi}}} \delta _{n_1,n_2}} \Bigg\} \ ,
\end{align}
with $x_{\pi}=\frac{k^+}{P^+}$. Note that we have $0<x_{\pi}<1$ in this work.

\subsection{Interaction terms for the $N\pi$ system}
Up to the level of the one-pion processes, the interaction terms in $P^-$ can be sorted into the pion-absorption term and the pion-emission term
\begin{eqnarray}
P^-_{\rm int} =&  P^-_{\rm int; abs} +  P^-_{\rm int; em} \ .
\end{eqnarray} 

For an incoming nucleon (labeled ``2") that absorbs a pion (carrying momentum $k$ and isospin projection $\lambda$) and the outgoing nucleon (labeled ``1"), the term corresponding to one-pion absorption is
\begin{align}
P^-_{\rm int; abs}	=& i M_N {\frac{g_A}{F} } \sum _{p_1^+} \sum _{p_2^+} \sum _{k^+} \frac{1}{2\pi\sqrt{2Lk^+}} \delta (p^+_1|k^+ + p^+_2) \nonumber \\
	& \sum _{s_1,s_2} \sum _{t_1,t_2} \sum _{\lambda} \int \frac{d^2p^{\perp}_1}{\sqrt{(2\pi)^2}} \frac{d^2k^{\perp}}{\sqrt{(2\pi)^2}} \frac{d^2p^{\perp}_2}{\sqrt{(2\pi)^2}}  \delta ^{(2)} (p^{\perp}_1-k^{\perp}-p^{\perp}_2) \nonumber \\
				& \times b^{\dagger} (p_1,s_1,t_1) a(k, \lambda) b(p_2,s_2,t_2)\nonumber \\ 
				& \times \underbrace{ \zeta ^{\dagger} (s_1) \Bigg\{ \frac{\gamma ^{\perp}\cdot p^{\perp}_1 +M_N}{p^+_1} \gamma _5 - \gamma _5 \frac{-\gamma ^{\perp}\cdot p^{\perp}_2 +M_N}{p^+_2} \Bigg\} \zeta (s_2) } _{\mbox{spinor kernel}} \ \underbrace{ T^{\dagger}(t_1) \Big[ \sum _a \tau _a \varepsilon _a (\lambda) \Big] T(t_2)}_{\mbox{isospinor kernel}} \ ,
\end{align}
where $\delta (p^+_1|k^+ + p^+_2)$ is the Kronecker delta for the discretized longitudinal momenta ($p^+_1$, $k^+$ and $p^+_2$), which ensures the conservation of the longitudinal momentum during the pion absorption. The spinor kernel for different helicity configurations of the incoming and outgoing nucleons is
\begin{align}
\begin{array}{|c|c|c|}
s_1 & \zeta ^{\dagger} (s_1) \Bigg\{ \frac{\gamma ^{\perp}\cdot p^{\perp}_1 +M_N}{p^+_1} \gamma _5 - \gamma _5 \frac{-\gamma ^{\perp}\cdot p^{\perp}_2 +M_N}{p^+_2} \Bigg\} \zeta (s_2) & s_2  \\ 
\uparrow & \frac{1}{p_1^+} M_N - \frac{1}{p_2^+} M_N & \uparrow \\ 
\uparrow & \frac{1}{p_1^+}(p_1^{\perp})^{\ast} - \frac{1}{p_2^+}(p_2^{\perp})^{\ast} & \downarrow \\ 
\downarrow & \frac{1}{p_1^+} p^{\perp}_1 - \frac{1}{p_2^+} p^{\perp}_2 & \uparrow \\ 
\downarrow & -\frac{1}{p_1^+} M_N + \frac{1}{p_2^+} M_N & \downarrow 
\end{array} \ \ .
\end{align} 
For clarity, we use ''$\uparrow $" and ''$\downarrow$" to denote the values of $+\frac{1}{2}$ and $- \frac{1}{2}$, respectively. The isospinor kernel for different isospin configurations of the incoming and outgoing nucleons is
\begin{align}
\begin{array}{|c|c|c||c|}
t_1 & T^{\dagger}(t_1) \Big[ \sum _a \tau _a \varepsilon _a (\lambda) \Big] T(t_2) & t_2 & \lambda =t_1 -t_2 \\ 
\uparrow & 1 & \uparrow & 0 \\ 
\uparrow & \sqrt{2} & \downarrow & 1 \\ 
\downarrow & \sqrt{2} & \uparrow & -1 \\ 
\downarrow & -1 & \downarrow & 0
\end{array} \ .
\end{align}
Applying Eqs. \eqref{eq:aToAlpha}, \eqref{eq:bToBeta} and Eq. \eqref{eq:MomentumFractionWeightedVariables}, we obtain the contribution from the one-pion absorption term to the mass squared operator [Eq. \eqref{eq:HLCoperator}]:
\begin{align}
 P^+  P^-_{\rm int; abs} 
=& i M_N { \frac{g_A}{F} } \frac{1}{\sqrt{4\pi K}} \sum _{x_1} \sum _{x_2} \sum _{x_k} \sum _{s_1, s_2} \sum _{t_1,t_2} \sum _{\lambda} \sum _{n_1,m_1} \sum _{n_2,m_2} \sum _{n_k,m_k} \ \sqrt{x_1 x_2} \  \delta (x_1 |x_k +x_2)  \nonumber \\
 & \times  \beta ^{\dagger} (x_1,n_1,m_1,s_1,t_1)\ \alpha (x_k,n_k,m_k,\lambda)\ \beta (x_2,n_2,m_2,s_2,t_2) \  T^{\dagger}(t_1) \Big[ \sum _a \tau _a \varepsilon _a (\lambda) \Big] T(t_2) \nonumber \\
	& \times  \int \frac{d^2 q_1^{\perp}}{(2\pi)^2}  \frac{d^2 q_k^{\perp}}{(2\pi)^2}  \frac{d^2 q_2^{\perp}}{(2\pi)^2} (2\pi)^2 \delta ^{(2)} (\sqrt{x_1}q_1^{\perp}-\sqrt{x_k}q_k^{\perp}-\sqrt{x_2}q_2^{\perp})  \nonumber \\
	&  \ \ \ \ \ \ \times \begin{cases}
		\  \ {\Psi _{n_1}^{m_1}}^{\ast}(q_1^{\perp}) \ \big[ \frac{M_N}{x_1} -\frac{M_N}{x_2} \big] \ \Psi _{n_k}^{m_k}(q^{\perp}_k) \Psi _{n_2}^{m_2}(q_2^{\perp}), \ \ \ \ \ \ \ \ \ \ \  \ \ \ \ \  \mbox{for} \  s_1 = \uparrow ,\ s_2 = \uparrow \\
		\ \ {\Psi _{n_1}^{m_1}}^{\ast}(q_1^{\perp}) \ \big[ \frac{1}{\sqrt{x_1}}(q_1^{\perp})^{\ast} - \frac{1}{\sqrt{x_2}}(q_2^{\perp})^{\ast}  \big] \ \Psi _{n_k}^{m_k}(q^{\perp}_k) \Psi _{n_2}^{m_2}(q_2^{\perp}), \ \mbox{for}  \ s_1 =\uparrow ,\  s_2 = \downarrow \\
		\ \ {\Psi _{n_1}^{m_1}}^{\ast}(q_1^{\perp}) \ \big[ \frac{1}{\sqrt{x_1}} q_1^{\perp} - \frac{1}{\sqrt{x_2}} q_2^{\perp} \big] \ \Psi _{n_k}^{m_k}(q^{\perp}_k) \Psi _{n_2}^{m_2}(q_2^{\perp}), \ \ \ \ \ \ \ \ \mbox{for} \ s_1 =\downarrow ,\ s_2 = \uparrow \\
		\ \ {\Psi _{n_1}^{m_1}}^{\ast}(q_1^{\perp}) \ \big[ -\frac{M_N}{x_1} + \frac{M_N}{x_2}  \big] \ \Psi _{n_k}^{m_k}(q^{\perp}_k) \Psi _{n_2}^{m_2}(q_2^{\perp}) \ \ \ \ \ \ \ \ \ \ \ \ \ \mbox{for} \ s_1 =\downarrow ,\  s_2 = \downarrow
		\end{cases} \ ,
\end{align}
where we have also substituted the identity $P^+ = \frac{2\pi}{L}K$. The longitudinal momentum fractions are $x_1=\frac{p_1^+}{P^+}$, $x_2=\frac{p_2^+}{P^+}$ and $x_k=\frac{p_k^+}{P^+}$. The analytic expression of the matrix element $P^+  \cdot P^-_{\rm int; abs}$ in the LF representation can be evaluated applying the identities in Appendix \ref{sec:2DHOIntegrals}.

Note that the one-pion emission contribution to the mass squared operator is the Hermitian conjugate of the one-pion absorption term $P^+P^-_{\rm int; abs}$.

\section{Proton's Dirac form factor}
\label{sec:DiracFFformula}
The Dirac form factor for the physical proton [\eqref{eq:formfactorExtended}] is 
\begin{align}
F_{1} (Q^2) =& F_{1,f} ^{p} (Q^2) + F_{1,f} ^{p \pi ^0} (Q^2) + F_{1,b} ^{n \pi ^+} (Q^2) \ \nonumber .
\end{align}
Note that $q^2$ is substituted by $Q^2$ according to Eq. \eqref{eq:squaredTransMomentumTransfer}.

The first contribution is
\begin{align}
F_{1,f} ^{p} (Q^2)=& {\sum _{t_N, n_N, m_N, s_N } } \ C^{\ast} ( x_N, n_N, m_N, s_N, t_N)  C ( x_N, n_N, m_N , s_N, t_N ) \ \label{eq:bareNDiracFormFactor} ,
\end{align}
which results from the virtual photon coupling to the current of the bare proton $|p \rangle$. Here, the basis quantum numbers (according to Eq. \eqref{eq:LihntFrontbasisSet}) are shown explicitly for clarity. The subscript ``$f$" denotes the contribution from the fermionic current. The summation in Eq. \eqref{eq:bareNDiracFormFactor} is only for the bare proton sector, i.e., $x_N=1$. In fact, $F_{1,f} ^{p} (Q^2)$ is the probability of the bare proton sector, $ |a_p|^2$ (according to Eq. \eqref{eq:fockSectorTruncationProton}), and it is independent of $Q^2$.

The second contribution is 
\begin{align}
F_{1,f} ^{p \pi ^0} (Q^2) =& \sum _{x_N} \sum _{s_N } \sum _{t_N , \lambda} \sum _{n_N',m_N'} \sum _{n_{\pi}',m_{\pi}'} \sum _{n_N,m_N} \sum _{n_{\pi},m_{\pi}}  e\big( t_N \big) \nonumber \\
& \ \ \ \ \ \ \ \ \ \ \ \times  C ^{\ast}( x_N , n'_N, m'_N , s_N, t_N; x_{\pi}, n'_{\pi} , m'_{\pi}  , \lambda )  C ( x_N, n_N, m_N , s_N, t_N; x_{\pi} , n_{\pi} , m_{\pi} , \lambda) \nonumber \\
& \ \ \ \ \ \ \ \ \ \ \ \times  \langle n'_N, m'_N ; \frac{x_{\pi}}{\sqrt{x_N}} q^{\perp} | n_N, m_N \rangle \ \langle n'_{\pi} , m'_{\pi} ; - \frac{x_{\pi}}{\sqrt{x_{\pi}}} q^{\perp} | n_{\pi} , m_{\pi} \rangle \ . 
\end{align}
$F_{1,f} ^{p \pi ^0} (Q^2)$ denotes the contribution from the virtual photon coupling to the current of the bare proton when dressed by charge-neutral $\pi ^0$. The effective charge factor of the nucleons is 
\begin{align}
e(t_N)=
\begin{cases}
\ 1 \ \ \ & {\rm for} \ \ t_N=+1/2 \\
\ 0 \ \ \ & {\rm for} \ \ t_N=-1/2 \ 
\end{cases} \ .
\end{align}
The kernel in the last line is the shifted operator, which is defined in Appendix \ref{sec:ShitedOperator}. This kernel, hence $F_{1,f} ^{p \pi ^0} (Q^2)$, vanishes as $Q^2 \rightarrow \infty $. At the limit of $Q^2=0$, $F_{1,f} ^{p \pi ^0} (0)= |a_{p \pi ^0}|^2$, which represents the probability of the $| p \pi ^0 \rangle$ sector [Eq. \eqref{eq:fockSectorTruncationProton}].

The third contribution is 
\begin{align}
F_{1,b} ^{n \pi ^+} (Q^2) =& \sum _{x_N} \sum _{s_N } \sum _{t_N , \lambda } \sum _{n_N',m_N'} \sum _{n_{\pi}',m_{\pi}'} \sum _{n_N,m_N} \sum _{n_{\pi},m_{\pi}}  e\big( \lambda \big)    \nonumber \\ 
& \ \ \ \ \ \ \ \ \ \ \ \times   C ^{\ast}( x_N, n'_N, m'_N, s_N , t_N; x_{\pi} , n'_{\pi} , m'_{\pi} , \lambda )  C ( x_N, n_N, m_N, s_N , t_N; x_{\pi} , n_{\pi} , m_{\pi} ,\lambda ) \nonumber \\
& \ \ \ \ \ \ \ \ \ \ \ \times \langle  n'_N, m'_N ;  - \frac{x_{N}}{\sqrt{x_N}} q^{\perp} |  n_N, m_N \rangle \ \langle  n'_{\pi} , m'_{\pi} ; \frac{x_{N}}{\sqrt{x_{\pi}}} q^{\perp} |  n_{\pi} , m_{\pi} \rangle \ .
\end{align}
$ F_{1,b} ^{n \pi ^+} (Q^2) $ denotes the contribution from the virtual photon coupling to the current of $\pi ^+$ that dresses the bare neutron. The subscript ``$b$" denotes the contribution from the bosonic current. The effective charge factor of the pions is 
\begin{align}
e\big( \lambda \big)=
\begin{cases}
\ +1 \ \ \ & {\rm for} \ \ \lambda=+1 \\
\ 0 \ \ \ & {\rm for} \ \ \lambda=0 \\
\ -1 \ \ \ & {\rm for} \ \ \lambda=-1 \
\end{cases} \ .
\end{align}
Analogous to $F_{1,f} ^{p \pi ^0} (Q^2)$, $F_{1,b} ^{n \pi ^+} (Q^2)$ vanishes for $Q^2 \rightarrow \infty $. At the limit of $Q^2=0$, $F_{1,b} ^{n \pi ^+} (0)= |a_{n \pi ^+}|^2$, which represents the probability of the $| n \pi ^+ \rangle$ sector [Eq. \eqref{eq:fockSectorTruncationProton}].

\section{Talmi-Moshinsky transformation}
The Talmi-Moshinsky (TM) transformation of the 2DHO wave function [Eq. \eqref{eq:2DHOinComplexMomentumRep}] is defined via the following relation:
\begin{align}
\Psi _{n_1}^{m_1} (q_1^{\perp}) \Psi _{n_2}^{m_2} (q_2^{\perp}) = \sum _{NMnm} \mathcal{M}_{n_1,m_1,n_2,m_2}^{N,M,n,m} (x_1,x_2) \Psi _N^M (Q^{\perp}) \Psi _n^m (q^{\perp}) \ ,
\end{align}
where the TM bracket is defined as
\begin{align}
\mathcal{M}_{n_1,m_1,n_2,m_2}^{N,M,n,m} (x_1,x_2) \equiv \langle NMnm | n_1 m_1n_2m_2 \rangle \ ,
\end{align}
with $2n_1+|m_1|+2n_2+|m_2|=2N+|M|+2n+|m|$ and $m_1+m_2=M+m$. The analytic expression of the TM bracket can be found in Refs. \cite{Li:2013cga,Wiecki:2014ola,Chaos:2004GoodLuck}. $q_1^{\perp}$ and $q_2^{\perp}$ are defined according to Eq. \eqref{eq:MomentumFractionWeightedVariables} as
\begin{align}
q_1^{\perp} = \frac{p^{\perp}_1}{\sqrt{x_1}}, \ q_2^{\perp} = \frac{p^{\perp}_2}{\sqrt{x_2}} \ .
\end{align}
The relative momentum $q^{\perp}$ and COM momentum $Q^{\perp}$ are, respectively,
\begin{align}
q^{\perp} =& \frac{\sqrt{x_2} q_1^{\perp} - \sqrt{x_1}  q_2^{\perp} }{\sqrt{x_1+x_2}} \ , \\
Q^{\perp} =& \frac{\sqrt{x_1} q_1^{\perp} + \sqrt{x_2}  q_2^{\perp} }{\sqrt{x_1+x_2}} \ .
\end{align}

\section{Some integrals involving the 2DHO basis}
\label{sec:2DHOIntegrals}

\subsection{Identities}
\begin{align}
\mathcal{P}^{(k)} (n,m) = \int \frac{d^2q^{\perp}}{(2\pi)^2} \big(q^{\perp} \big)^k \Psi _n^m (q^{\perp}) = {\mathit{b}}^{k+1} (-1)^n 2^k \sqrt{\frac{(n+k)!}{\pi n!}} \delta _{k,-m} \ .
\end{align}

\begin{align}
\mathcal{PC}^{(k)} (n,m) = \int \frac{d^2q^{\perp}}{(2\pi)^2} \Big[\Big(q^{\perp}\Big)^{\ast} \Big]^k \Psi _n^m (q^{\perp}) = {\mathit{b}}^{k+1} (-1)^n 2^k \sqrt{\frac{(n+k)!}{\pi n!}} \delta _{k,m} \ .
\end{align}

\begin{align}
\mathcal{P}^{(1)}(n',m';n,m) = & \langle  n' m' | q^{\perp} | n m \rangle \ = \ \langle n'm' | q^{\perp} |nm \rangle \nonumber \\
=&  \mathit{ b} \ \delta _{m', m+1} 
	\begin{cases}
	\ \sqrt{n+|m|+1} \delta _{n,n'} - \sqrt{n} \delta _{n,n'+1} \ , \ \ m \geq 0 , \ n \geq n' \\
	\ \sqrt{n+|m|} \delta _{n,n'} - \sqrt{n+1} \delta _{n', n+1} \ , \ \ m<0 , \ n \leq n'
	\end{cases} \ .
\end{align}

\begin{align}
\mathcal{PC}^{(1)}(n',m';n,m) = & \langle  n' m' | \big( q^{\perp} \big)^{\ast} | n m \rangle \ = \  \langle n'm' | \big( q^{\perp} \big)^{\ast} |nm \rangle \nonumber \\
=&  \mathit{ b} \ \delta _{m,m'+1} 
	\begin{cases}
	\ \sqrt{n'+|m'|+1} \delta _{n,n'} - \sqrt{n'} \delta _{n',n+1} \ , \ \ \ m' \geq 0 \\
	\ \sqrt{n' +|m'|} \delta _{n,n'} - \sqrt{n'+1} \delta _{n,n'+1} \ , \ \ \ m' <0
	\end{cases} \ .
\end{align}

\begin{align}
 & \langle n'm' | q^{\perp} (q^{\perp})^{\ast} | nm \rangle \ = \ \int \frac{d^2 q^{\perp}}{(2\pi)^2} \Big( \Psi _{n'}^{m'} (q^{\perp}) \Big)^{\ast} |q^{\perp}|^2 \Psi _{n}^{m} (q^{\perp}) \nonumber \\
=& b^2 \delta _{m',m} \Big[ (2n + |m| +1) \delta _{n',n} -\sqrt{n'(n'+|m'|)} \delta _{n',n+1} - \sqrt{n(n+|m|)} \delta _{n,n'+1} \Big] \label{eq:KEterm2DHOintegral} \ .
\end{align}

\subsection{Shifted operator}
\label{sec:ShitedOperator}
The shifted operator, in the 2DHO representation, is defined as $\langle n',m'; u^{\perp} + q^{\perp} | n,m; u^{\perp} \rangle$, where the initial and final transverse momenta are centered at $u^{\perp}$ and $u^{\perp} + q^{\perp}$, respectively. According to the translational invariance of the 2DHO basis function, it can be evaluated as
\begin{align}
\langle n',m'; u^{\perp} +q^{\perp} | n,m; u^{\perp} \rangle =& \langle n',m'; u^{\perp} + \frac{1}{2} q^{\perp} | n,m; u^{\perp} -\frac{1}{2} q^{\perp} \rangle \ .
\end{align}
Applying the 2DHO wave function in the complex momentum representation [Eq. \eqref{eq:2DHOinComplexMomentumRep}], the shifted operator reads 
\begin{align}
\int \frac{d^2 u^{\perp}}{(2\pi)^2} \Big( \Psi _{n'}^{m'} (u^{\perp} +q^{\perp}) \Big)^{\ast}  \Psi _{n}^m (u^{\perp})  = &  \int \frac{d^2 u^{\perp}}{(2\pi)^2} {\Psi  _{n'}^{m'}}^{\ast} (u^{\perp} + \frac{1}{2} q^{\perp}) \Psi _{n}^m (u^{\perp} -\frac{1}{2} q^{\perp} ) \nonumber \\
=& \sum _{\nu} \mathcal{M} _{n',-m',n,m} ^{N,0,\nu , \mu} (\frac{\pi}{4})\  \frac{b}{\sqrt{4\pi}} (-1)^N \Psi _{\nu } ^{\mu} \big( \frac{1}{\sqrt{2}} q^{\perp} \big) \ ,
\end{align}
with 
\begin{align}
\mu =& m - m' , \\ 
N =& n' +n -\nu + \frac{1}{2} \big( |m'| +|m| - |\mu| \big) \ , \\
0 \leq & \nu \leq n +n' + \frac{1}{2} \big(  |m'| +|m| - |\mu| \big) \ .
\end{align}

\subsection{Integrals involving three 2DHO basis functions}
\begin{align}
& \int \frac{d^2 q_1^{\perp}}{(2 \pi)^2} \frac{d^2 q_2^{\perp}}{(2 \pi)^2} \frac{d^2 {q'}^{\perp}}{(2 \pi)^2} (2\pi)^2 \delta ^2 (\sqrt{x_1} q_1^{\perp} +\sqrt{x_2} q_2^{\perp} -\sqrt{x'} {q'}^{\perp} ) \ \Psi _{n_1}^{m_1}(q_1^{\perp}) \Psi _{n_2}^{m_2} (q_2^{\perp}) {\Psi _{n'}^{m'}}^{\ast} ({q'}^{\perp}) \nonumber \\
=& \delta _{m_1+m_2,m'}\ \frac{1}{x'}\ \mathcal{M}_{n_1,m_1,n_2,m_2}^{n',m',n,0}(x_1,x_2)\ \mathcal{P}^{(0)}(n,0) \ ,
\end{align}
where $n=n_1 +n_2 -n' + \frac{1}{2}(|m_1|+|m_2|-|m_1+m_2|) \geq 0$. 

\begin{align}
& \int \frac{d^2 q_1^{\perp}}{(2 \pi)^2} \frac{d^2 q_2^{\perp}}{(2 \pi)^2} \frac{d^2 {q'}^{\perp}}{(2 \pi)^2} (2\pi)^2 \delta ^2 (\sqrt{x_1} q_1^{\perp} +\sqrt{x_2} q_2^{\perp} -\sqrt{x'} {q'}^{\perp} ) \cdot {q'}^{\perp} \cdot \Psi _{n_1}^{m_1}(q_1^{\perp}) \Psi _{n_2}^{m_2} (q_2^{\perp}) {\Psi _{n'}^{m'}}^{\ast} ({q'}^{\perp}) \nonumber \\
=&  \delta _{m_1+m_2,m'-1} \ \frac{1}{x'} \sum _{N=max[0,n'-1]}^{min[\nu , n'+1]} \mathcal{M} _{n_1,m_1,n_2,m_2} ^{N,m'-1, \nu -N , 0}(x_1,x_2) \ \mathcal{P}^{(1)}(n',m';N,m'-1) \ \mathcal{P}^{(0)} (\nu -N,0) \,
\end{align}
where $\nu =N+n = n_1+n_2+\frac{1}{2}(|m_1|+|m_2|-|m_1+m_2|)$.

\begin{align}
& \int \frac{d^2 q_1^{\perp}}{(2 \pi)^2} \frac{d^2 q_2^{\perp}}{(2 \pi)^2} \frac{d^2 {q'}^{\perp}}{(2 \pi)^2} (2\pi)^2 \delta ^2 (\sqrt{x_1} q_1^{\perp} +\sqrt{x_2} q_2^{\perp} -\sqrt{x'} {q'}^{\perp} ) \cdot ({q'}^{\perp})^{\ast} \cdot \Psi _{n_1}^{m_1}(q_1^{\perp}) \Psi _{n_2}^{m_2} (q_2^{\perp}) {\Psi _{n'}^{m'}}^{\ast} ({q'}^{\perp}) \nonumber \\
=& \delta _{m_1+m_2,m'+1} \ \frac{1}{x'} \sum _{N=max[0,n'-1]}^{min[\nu , n'+1]} \mathcal{M}_{n_1,m_1,n_2,m_2}^{N,m'+1,\nu -N,0}(x_1,x_2) \ \mathcal{PC}^{(1)}(n',m';N,m'+1)\ \mathcal{P} ^{(0)}(\nu -N,0) \ ,
\end{align}
where $\nu =N+n=n_1+n_2 + \frac{1}{2}(|m_1|+|m_2|-|m'+1|)$.

\begin{align}
& \int \frac{d^2 q_1^{\perp}}{(2 \pi)^2} \frac{d^2 q_2^{\perp}}{(2 \pi)^2} \frac{d^2 {q'}^{\perp}}{(2 \pi)^2} (2\pi)^2 \delta ^2 (\sqrt{x_1} q_1^{\perp} +\sqrt{x_2} q_2^{\perp} -\sqrt{x'} {q'}^{\perp} ) \cdot {q}_1^{\perp} \cdot \Psi _{n_1}^{m_1}(q_1^{\perp}) \Psi _{n_2}^{m_2} (q_2^{\perp}) {\Psi _{n'}^{m'}}^{\ast} ({q'}^{\perp}) \nonumber \\
=& \delta _{m_1+m_2,m'-1} \Bigg\{ \sqrt{\frac{x_1}{(x_1+x_2)^3}}\sum _{N=max[0,n'-1]}^{min[n'+1,\nu]} \mathcal{M} _{n_1,m_1,n_2,m_2}^{N,m'-1,\nu -N,0}(x_1,x_2) \ \mathcal{P}^{(1)}(n',m';N,m'-1) \ \mathcal{P}^{(0)}(\nu -N ,0) \nonumber \\
 & \ \ \ \ \ \ \ \ \ \ \ \ \ \ \ \ \ \ \ + \sqrt{\frac{x_2}{(x_1+x_2)^3}}\ \theta (n)\ \mathcal{M}_{n_1,m_1,n_2,m_2}^{n',m',n,-1}(x_1,x_2) \ \mathcal{P} ^{(1)} (n,-1) \Bigg\} \ ,
\end{align}
where $ \nu = N + n = n_1 + n_2 + \frac{1}{2}(|m_1|+|m_2|-|m'-1|)$ and $n=n_1+n_2-n' + \frac{1}{2}(|m_1|+|m_2|-|m'|-1) \geq 0 $ .

\begin{align}
& \int \frac{d^2 q_1^{\perp}}{(2 \pi)^2} \frac{d^2 q_2^{\perp}}{(2 \pi)^2} \frac{d^2 {q'}^{\perp}}{(2 \pi)^2} (2\pi)^2 \delta ^2 (\sqrt{x_1} q_1^{\perp} +\sqrt{x_2} q_2^{\perp} -\sqrt{x'} {q'}^{\perp} ) \cdot \big( {q}_1^{\perp} \big)^{\ast} \cdot \Psi _{n_1}^{m_1}(q_1^{\perp}) \Psi _{n_2}^{m_2} (q_2^{\perp}) {\Psi _{n'}^{m'}}^{\ast} ({q'}^{\perp}) \nonumber \\
=& \delta _{m_1+m_2,m'+1} \Bigg\{ \sqrt{\frac{x_1}{(x_1+x_2)^3}} \sum _{N=max[0,n'-1]}^{min[n'+1,\nu]} \mathcal{M} _{n_1,m_1,n_2,m_2}^{N,m'+1,\nu -N,0}(x_1,x_2) \ \mathcal{PC} ^{(1)} (n',m'; N , m'+1) \mathcal{P} ^{(0)} (\nu -N , 0) \nonumber \\
 & \ \ \ \ \ \ \ \ \ \ \ \ \ \ \ \ \ \ \ + \sqrt{\frac{x_2}{(x_1+x_2)^3}} \ \theta (n)\  \mathcal{M} _{n_1,m_1,n_2,m_2}^{n',m',n,1}(x_1,x_2) \ \mathcal{PC} ^{(1)} (n,1) \ \Bigg\} \ ,
\end{align}
where $ \nu = N+n = n_1 +n_2 + \frac{1}{2}(|m_1| +|m_2| - |m'+1| ) $ and $ n= n_1 + n_2 - n' + \frac{1}{2} (|m_1| + |m_2| -|m'|-1) \geq 0 $.

\end{appendices}


\end{document}